\documentclass[9pt, twocolumn]{extarticle}
\usepackage{graphicx}
\usepackage{amsmath}
\usepackage{amssymb}
\usepackage[gen]{eurosym}
\usepackage{xcolor}
\usepackage{tabularx}
\usepackage{authblk}

\usepackage{geometry}
\definecolor{asalinkcolor}{cmyk}{1,.5,0,0}
\usepackage[bookmarks=false,         
     pdfnewwindow=true,      
     colorlinks=true,    
     linkcolor=asalinkcolor,     
     citecolor=asalinkcolor,     
     filecolor=asalinkcolor,  
     urlcolor=asalinkcolor,      
    final=true]{hyperref}

\newcommand{\subix}[2][]{_{\text{#2}#1}}
\newcommand{\supix}[2][]{^{\text{#2}#1}}
\newcommand{\AMPA}{\supix{AMPA}}
\newcommand{\GABA}{\supix{GABA}}
\newcommand{\NMDA}{\supix{NMDA}}

\begin{document}

  \title{Neural modelling of the encoding of fast frequency modulation}
  \author[12]{Alejandro Tabas}
  \author[2]{Katharina von Kriegstein}
  \affil[1]{Chair of Cognitive and Clinical Neuroscience, Faculty of Psychology, Technische Universit\"{a}t Dresden, 01062 Dresden, Saxony, Germany}
  \affil[2]{Max Planck Institute for Human Cognitive and Brain Sciences, Stephanstr 1a, 04107 Leipzig, Saxony, Germany}
  \date{}

\onecolumn

  \maketitle

  \begin{abstract} 

    Frequency modulation (FM) is a basic constituent of vocalisation in many animals as well as in humans. In human speech, short rising and falling FM-sweeps called formant transitions characterise individual speech sounds. There are two representations of FM in the ascending auditory pathway: a spectral representation, holding the instantaneous frequency of the stimuli; and a sweep representation, consisting of neurons that respond selectively to FM direction. To-date computational models use feedforward mechanisms to explain FM encoding. However, from neuroanatomy we know that there are massive feedback projections in the auditory pathway. Here, we found that a classical FM-sweep perceptual effect, the sweep pitch shift, cannot be explained by standard feedforward processing models.  We hypothesised that the sweep pitch shift is caused by a predictive interaction between the sweep and the spectral representation. To test this hypothesis, we developed a novel model of FM encoding incorporating a predictive feedback mechanism. The model fully accounted for experimental data that we acquired in a perceptual experiment with human participants as well as previously published experimental results. We  also designed a new family of stimuli for a second perceptual experiment to further validate the model. Combined, our results indicate that predictive interaction between different frequency encoding and direction encoding neural representations plays an important role in the neural processing of FM. In the brain, this mechanism is likely to occur at early stages of the processing hierarchy. 

  \end{abstract}
  
\twocolumn

  \section{Introduction}
 
    Frequency modulation (FM) is a basic acoustic feature of animal vocalisation, human speech and music. In human speech, consonants preceding and following a vowel can be acoustically characterised by formant transitions: a series of simultaneous fast FM sinusoids of around 50\,ms duration that start or finish in the frequencies characterising the vowel \cite{Kent2008}. At all stages of the ascending auditory pathway, FM is encoded along the tonotopic axis in a \emph{spectral representation} that holds the instantaneous frequency of the stimuli \cite{Hu2003}. Individual neurons at higher levels of the processing hierarchy (inferior colliculus \cite{Geis2013, Li2010, Hage2003}, medial geniculate body \cite{Kuo2012, Lui2003}, and auditory cortex \cite{Issa2016, Trujillo2013, Ye2010, Zhang2003}) also encode FM direction and rate. We call this latter, more abstract representation, the \emph{sweep representation}.

    Despite the massive feedback projections that characterise the auditory pathway \cite{Schofield2011}, computational models to date use only feedforward mechanisms to explain FM encoding \cite{Skorheim2014}. Given the importance of high-order predictive elements in the optimisation of speech recognition abilities (e.g., \cite{Moore1995}), descending projections are likely to play an important role for how fast FM-sweeps, the basic building blocks of speech, are encoded in the auditory system. Feedback connections have general complex repercussions on the way sounds are processed by, for instance, modulating the properties of the receptive fields \cite{Shamma2014, Suga2012}.

    The sweep pitch shift is a classical behavioural effect from psychoacoustics first reported around 60 years ago \cite{Brady1961}. In the original experiment, participants listened to fast rising and falling FM-sweeps. The authors discovered that the participants judged up sweeps as eliciting a higher pitch than down sweeps with the same average fundamental frequency. These findings were later replicated \cite{Nabelek1970, Rossi1978}. To explain the effect d'Alessandro and colleagues proposed a phenomenological model assuming that the pitch of a sweep is integrated using a fixed-size window from the instantaneous frequency of the stimulus across time  \cite{DAlessandro1994, DAlessandro1998}. Due to the leaking memory of the integration, this process naturally favours the latest frequencies of the sweep, explaining the perceptual pitch shift. However, the authors found that different integration weights were necessary to explain different partitions of their data, indicating that the phenomenological model is not a parsimonious explanation of the sweep pitch shift. Whether classical mechanistic models of pitch processing (see~\cite{DeCheveigne2005} for a review) can explain the effect has not been considered before.

    Here we first showed that the sweep pitch shift \cite{Brady1961, Nabelek1970, Rossi1978} cannot be explained using classical models of pitch processing (see~\cite{DeCheveigne2005} for a review). We argue that the inability of these models to reproduce the perceptual results stems from the fact that they consist of feed-forward elements only, so that the spectral representation cannot interact with information related to the FM rate or direction. Similarly, since previous models of FM encoding \cite{Skorheim2014} considered a static representation of spectral information, they predict that the sweep pitch shift would not occur. The aim of this study is to build a comprehensive model of FM encoding incorporating both, the sweep and the spectral representations, and describing the sweep-to-spectral feedback mechanisms active during the processing of FM sounds.

    We approached this problem in three steps. First, we reexamined and quantified the sweep pitch shift in a behavioural experiment, and tested whether the experimental data could be explained by existing computational models of FM-encoding and pitch perception \cite{Zilany2014, Meddis1997, Meddis2006}. In the second step, we built a hierarchical model motivated by the hypothesis that the pitch sweep shift results from feedback modulation between the two representations. The feedforward components of the model were based on results of previous studies on FM direction selectivity and included frequency and FM sweep direction processing \cite{Skorheim2014, Ye2010, Razak2008}. The top-down architecture was grounded in the basis of generative hierarchical models and predictive coding \cite{Mumford1992, Friston2005} and informed by the human psychophysics results from the first part of the study. In the third and last step, we used a new set of stimuli termed \emph{sweep trains} to further validate the model. These stimuli, consisting of a concatenation of five sweeps, preserve the same acoustical features of the original sweeps but elicited different dynamics in the feedback system of the model than their single-sweep counterpart. The ability of the model to predict the pitch elicited by these novel stimuli illustrated the generalisation power of the computational mechanisms proposed in this work.

  \section{Results} 

    \subsection{The sweep pitch shift revisited}

      For the first behavioural study we used a total of $10 \times 3 = 30$ fast FM sweeps. The sweeps had 10 linearly distributed frequency spans $\Delta f \in [-600, 600]$\,Hz and 3 average frequencies $\bar{f} \in \{900, 1200, 1500\}$\,Hz. Each sweep had a duration of 40\,ms and was preceded and followed by 5\,ms segments of constant frequency. 8 participants matched the stimuli against probe pure tones of adjustable frequency. In one part of the experiment probe tones were presented before the sweep, in the other after the sweep. Each participant matched each stimulus four times. The pitch sweep shift was measured as the difference between the perceived pitch and the average frequency of the sweep: $\Delta p = f\subix{perceived} - \bar{f}$. Stimuli are available in the supporting information (\nameref{sounds}). 

\begin{figure*}
        \centering
        \includegraphics[width=\textwidth]{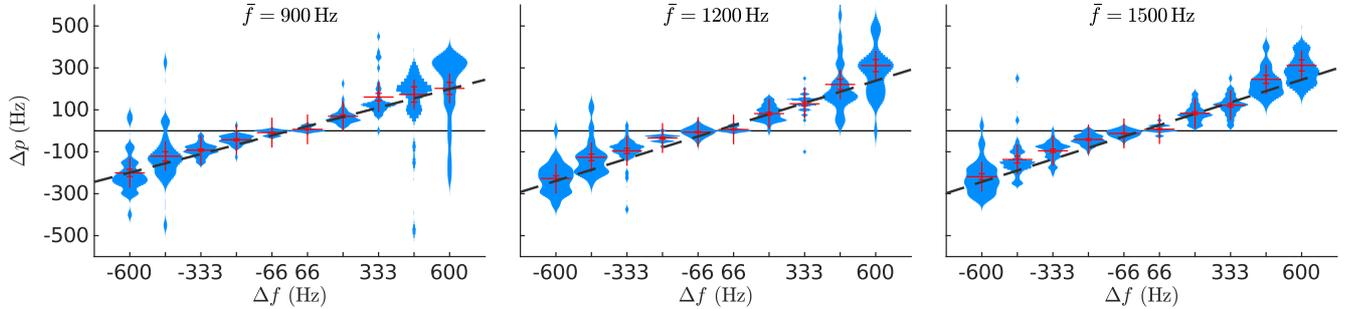}
        \caption{\textbf{Sweep pitch shift.} Kernel density estimations on the perceived pitch are plotted separately for each of the 30 sweeps used in the experiment. The $y$-axis of each plot shows the magnitude of the sweep pitch shift $\Delta p$. The x-axis list the gaps of each of the sweeps. Red crosses show the mean and standard error of the data. Dark dashed lines show the group linear fit of the data.} 
        \label{fig:swPitch}
      \end{figure*}

\begin{figure*}
        \centering
        \includegraphics[width=0.7\textwidth]{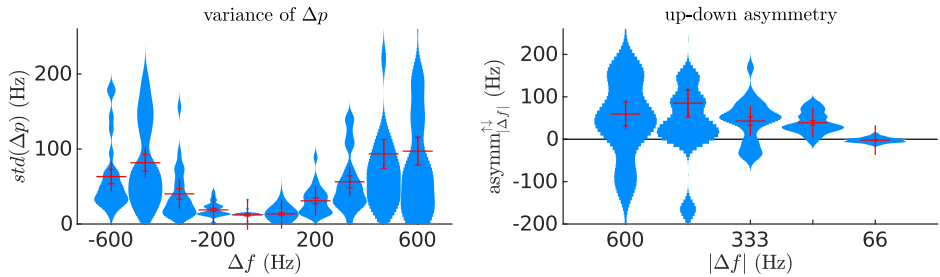}
        \caption{\textbf{Variance of the perceived pitch and up/down asymmetry.} Left: Kernel density estimations of the intra-subject standard deviation of the sweep pitch shift $\Delta p$, plotted separately for the different frequency spans $\Delta f$. Each sample in the distributions corresponds to the standard deviation of the perceived pitch of a sweep in one subject (i.e., in each distribution there are $8 \times 3 $ points, one for each subject and $\bar{f}$). The variance is monotonically correlated to the absolute gap $|\Delta f|$ ($r_s = 0.63$, $p < 10^{-27}$). Right: Kernel density estimations of the up/down asymmetry distributions as defined in Eq~\eqref{eq:udAsymm}. Each sample of the distributions corresponds to the difference of the average absolute deviation from centre frequency between up and down sweeps of the same $|\Delta f|$ for a given subject and centre frequency ($N = 8\times3 = 24$). Red crosses show the mean and the standard error of the data.} 
        \label{fig:swVar}
      \end{figure*}

      We found that the pitch shift $\Delta p$ depended on the sweep's span $\Delta f$ (Fig~\ref{fig:swPitch} and Table~Tab~\ref{tab:stats}). The exact dependence was consistent across listeners for sweeps with $\Delta f \leq 333$\,Hz lying in the vicinity of the linear fit $f_{\text{perceived}} \simeq \bar{f} + m\,\Delta f$ (with an average deviance from the fit of 46\,Hz). Sweeps with larger frequency spans resulted in wider distributions of $f_{\text{perceived}}$ due to higher inter- and intra- subject variabilities (Fig~\ref{fig:swVar}; see also~Fig.~\ref{fig:slopes}). Presenting the sweep before or after the probe tone did not systematically affect the perceived pitch (Fig.~\ref{fig:lrDev}).

      In their classical study, Brady and colleagues \cite{Brady1961} showed that the absolute value of the sweep pitch shift $|\Delta p|$ is larger for down than for up sweeps. In a later study, Nabelek and colleagues \cite{Nabelek1970} showed the reversed effect. To test if our data replicates any of these previous findings we drew, for each absolute frequency span $|\Delta f|$, the distribution of the differences between the pitch shift in up and down sweeps: 

      \begin{equation}
       \text{asymm}^{\uparrow \downarrow}_{|\Delta f|} = 
                                  |\Delta p(\Delta f)| - |\Delta p(-\Delta f)|
        \label{eq:udAsymm}
      \end{equation}

      Our results robustly replicated the observations from Nabelek and colleagues (Fig~\ref{fig:swVar}, right). The sweep pitch shift was significantly larger for up than down sweeps for \mbox{$|\Delta f| \leq 200\,$Hz} ($p < 2 \times 10^{-5}$) but not for \mbox{$|\Delta f| = 66\,$Hz} ($p = 0.77$), according to two-tailed rank-sum tests ($N = 96$). Up sweeps have been consistently found to be easier to discriminate from pure tones than down sweeps in a wide range of experimental conditions \cite{Luo2006b, Gordon2002, Madden1997, Collins1978}, probably because auditory nerve responses to up sweeps compensate for the low-frequency processing delay of the basilar membrane \cite{Uppenkamp2001}, provoking stronger neural responses than their down counterparts. The stronger pitch shift to up sweeps already suggest that the sweep representation plays an important role on the genesis of the sweep pitch shift. 

      Last, we tested if the dependence of the sweep pitch with $\Delta f$ was robustly replicated across subjects. The slope of the linear fit between $f_{\text{perceived}}$ and $\Delta f$, similar in magnitude in all participants, are plotted in Fig.~\ref{fig:slopes}.

    \subsection{Bottom-up models of pitch cannot explain the pitch shift}

      Pitch is represented in two complimentary codes within the auditory system: the spectral code, produced by the spectral decomposition of the stimuli performed by the basilar membrane; and the temporal code, comprised in the spike timings of the neurons across the auditory nerve that are phase locked to the stimulus waveform (see~\cite{Oxenham2013} for a review). If the sweep pitch shift was a consequence of bottom-up pitch processing, we would expect the effect to be explainable by previous computational models that use either of the two representations to infer pitch. To test this we computed the pitch predicted by one representative model of each family; i.e., one model using the spectral and one model using the temporal codes.

\begin{figure*}
        \centering
        \includegraphics[width=\textwidth]{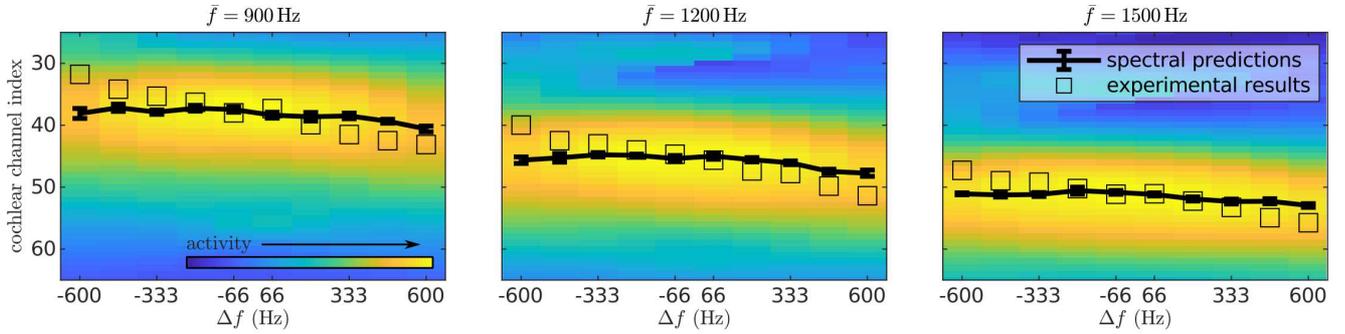}
        \caption{\textbf{Predictions of the spectral model of pitch.} Heatmaps show the mean activation across the duration of the stimuli at different cochlear channels, as simulated by a model of the auditory periphery in response to each sweep; units are arbitrary. Error bars point to the expected value and variance of the distribution across frequencies for each sweep. Each empty square denotes the expected channel elicited by a pure tone with the frequency of the average experimental data of the corresponding sweep.}
        \label{fig:spectralModel}
      \end{figure*}

      In the spectral model, pitch can be directly inferred by computing the expected value of the activity across cochlear channels in the auditory nerve \cite{DeCheveigne2005, Zilany2014}. Unlike the empirical data, predictions of the spectral model show no systematic dependence of $f_{\text{perceived}}$ on $\Delta f$ (Fig~\ref{fig:spectralModel}). Note that, since the sinusoidal FM-sweeps used in the experiments evoke a single peak in the spectral distributions, more sophisticated spectral models designed to explain how the pitch of harmonic complex tones would yield identical results \cite{DeCheveigne2005}. 

\begin{figure*}
        \centering
        \includegraphics[width=\textwidth]{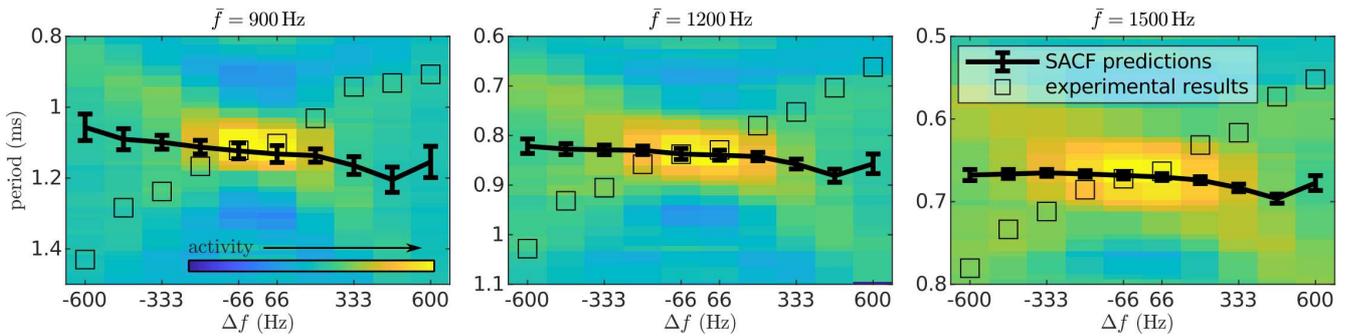}
        \caption{\textbf{Predictions of the temporal model of pitch.} Heatmaps show the distribution across periods elicited in the summary autocorrelation factor (SACF) for each sweep. The value corresponding to each period was computed as the average activation of the SACF across four harmonics (see Methods); units are arbitrary. Error bars point to the expected value and variance of the distribution across periods for each sweep. Each empty square denotes the expected period elicited in the SACF by a pure tone with the frequency of the average experimental data of the corresponding sweep.}
        \label{fig:sacfModel}
      \end{figure*}

      The temporal model was based on the principles of the summary autocorrelation function (SACF), that measures pitch according to the phase-locked response in the auditory nerve \cite{Meddis1997, Balaguer2008}. We chose this model because it performs a relatively straightforward analysis of the phase-locked activity in the periphery. Predictions of the temporal model lay within $f_{\text{perceived}} \simeq \bar{f}$ independently of $\Delta f$ (Fig.~\ref{fig:sacfModel}). This is most likely a consequence of the SACF being unable to decode rapidly changing frequencies with such short stimuli. 

    \subsection{The FM-feedback spectral model}

\begin{figure}
          \centering
          \includegraphics[width=\columnwidth]{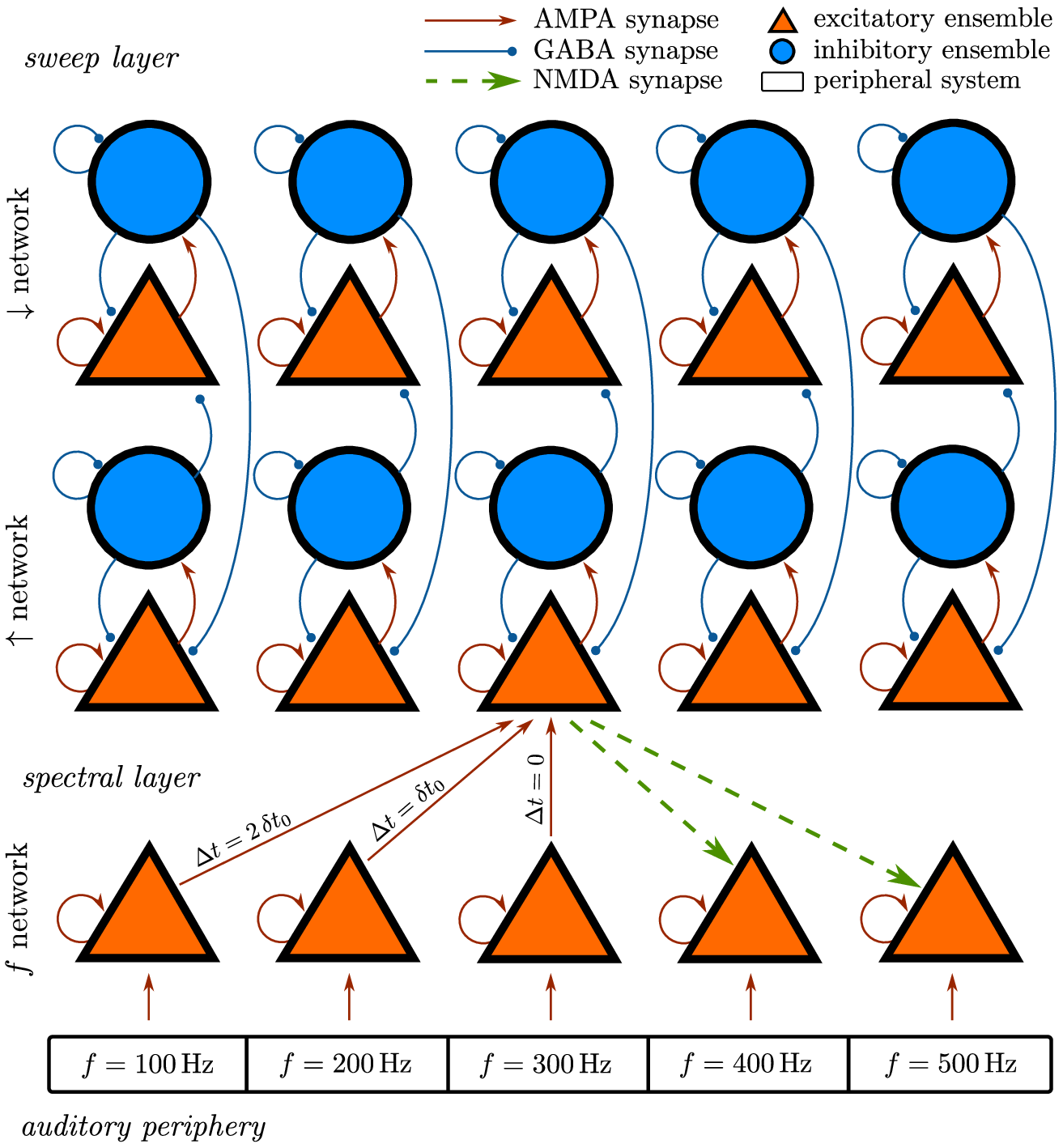}
          \caption{\textbf{Diagram of the FM-feedback spectral model.} The model consists of three layers: first, the \emph{peripheral system}, representing the activity at the beginning of the auditory nerve; second: the \emph{spectral layer}, with a network integrating the spectral information of the sound ($f$ network); and third, the \emph{sweep layer}, with one network specialised in detecting up ($\uparrow$ network) sweeps and another network specialised in detecting down ($\downarrow$ network) sweeps. The spectral layer integrates afferent inputs from the periphery and holds a representation of the stimulus that can be used to infer pitch. The sweep layer receives afferent inputs from the spectral layer that are used to decode the direction of the sweeps. Feedback connections from the sweep layer to the spectral layer modulate the time constants of the populations that are expected to be activated once the direction of the sweep has been decoded. The inhibitory ensembles in the up and down network enforce competition between up and down ensembles in a winner-take-all fashion. Note that the diagram is schematic and shows only 5 of the $N = 100$ populations and a single example of the connections between the sweep and the spectral layers. The labels of the boxes of the peripheral system are also schematic: the spectral resolution of the peripheral system is much higher.} 
          \label{fig:diagram}
        \end{figure}

        In this section we introduce a hierarchical model of FM-encoding, termed \emph{FM-feedback spectral model}, with two levels (Fig~\ref{fig:diagram}). In the first level, the \emph{spectral} layer holds a spectral representation of the sound. In the second level, the \emph{sweep} layer encodes FM-sweep direction. The spectral layer uses the spectral rather than the temporal code to represent the instantaneous frequency of the stimuli because, as we showed in the previous section, the phase-locked responses of the auditory peripheral model \cite{Zilany2014} are not able to catch up with the fast modulation of the frequency of the sweeps robustly. Moreover, the animal literature converges in the notion that sweep direction and rate are decoded from the spectral, and not the temporal representation of the sounds \cite{Skorheim2014, Kuo2012, Pollak2011, Li2010, Zhang2003, Lui2003}.
          
        The main hypothesis introduced in the FM-feedback spectral model model is that, once the direction of the sweep is encoded in the sweep layer, a feedback mechanism modulates the effective time constant of the populations encoding the frequencies that are expected to be activated next in the spectral layer. We expect this parsimonious mechanism to qualitatively explain why the posterior parts of the sweep are given a higher weight during perceptual integration and to quantitatively reproduce the exact dependence of pitch with $\Delta f$ observed in our data. An implementation of the FM-feedback spectral model written in python is freely available at \url{https://github.com/} (the libraries will be available there upon publication; the code is currently attached to the manuscript for revision).

\begin{figure*}
          \centering
          \includegraphics[width=0.7\textwidth]{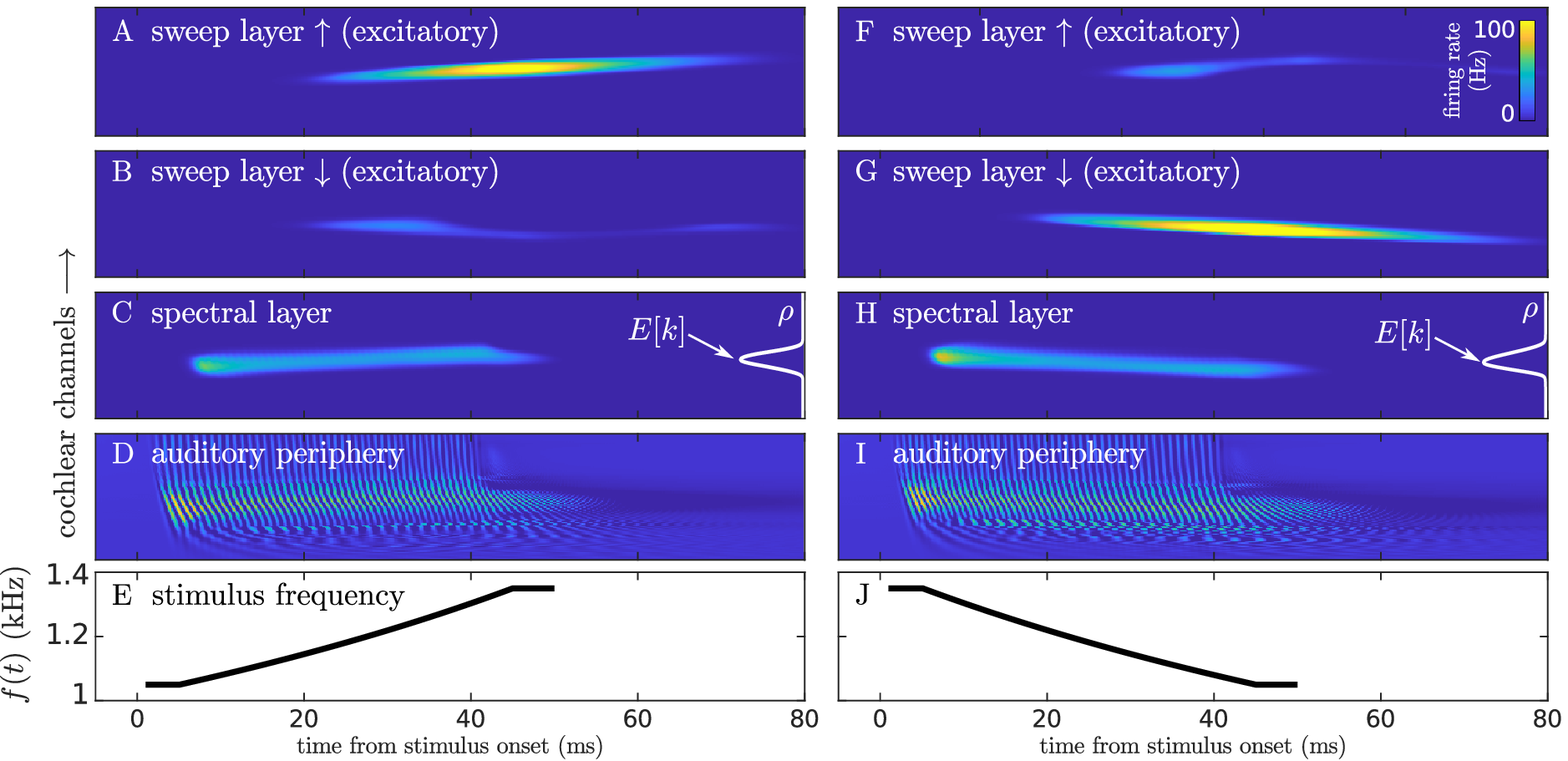}
          \caption{\textbf{Model ­responses to an up and a down sweep.} A-E show the responses to an up sweep, and F-J to a down sweep. From top to bottom: (A/F) the instantaneous firing rate of the up-selective excitatory populations in the sweep layer; (B/G) the instantaneous firing rate of the down-selective excitatory populations in the sweep layer; (C/H) the instantaneous firing rate of the populations in the spectral layer and, in the right of the panels, a schematic view of the probability distribution of pitch derived from this representation; (D/I) the output of the model of the auditory periphery; (E/F) the instantaneous frequency of the sweeps along time. In all panels except for E/J, $y$-axis represents the cochlear channel $n$, ordered from bottom to top. The stimuli were the up and a down sweeps with $\Delta = \pm300\,\text{Hz}$ and $\bar{f} = 1200\,\text{Hz}$ used in the experiment. 
          \label{fig:popResponses}}
        \end{figure*}

        Example responses of the excitatory populations of the model to up and down sweeps are shown in Fig~\ref{fig:popResponses}. 

      \subsubsection{Modelling FM direction selectivity \label{sec:mod:model:dsi}}
        
        At least three mechanisms for FM direction selectivity have been identified in the animal literature: asymmetric sideband inhibition \cite{Geis2013, Williams2012, Fuzessery2011}, duration sensitivity \cite{Morrison2018, Kuo2012, Williams2012}, and delayed excitation \cite{Fuzessery2011, Ye2010, Razak2008}. In order to prevent an excessive inflation in the dimensionality of the model's parameter space, we focus here on delayed excitation, a straightforward mechanism where neurons with different best frequencies output to the direction selective neuron with different delays; e.g., an up-selective neuron will receive delayed inputs from a neuron tuned to low frequencies and instantaneous inputs from a neuron tuned to high frequencies, so that an up sweep results in simultaneous excitation from both of them. Any related mechanism showing FM direction selectivity should yield similar overall results \cite{Skorheim2014}.

        In the FM-feedback spectral model, delayed excitation is implemented by introducing consistent delays between the populations in the spectral and the sweep layers. A sweep population receiving direct input from the spectral population encoding $f_0$ and responding selectively to up sweeps will receive increasingly delayed inputs from the spectral populations centred at $f < f_0$ (Fig~\ref{fig:diagram}). The relative delay in the connection between a spectral population $m$ and a target sweep population $n$ depends linearly on the spectral distance between the two ensembles: $\delta t_{nm} = |n-m| \delta t_0$.

        The sweep layer consists of two networks, each encoding one of the FM directions and responding selectively to \emph{up} ($\uparrow$) and \emph{down} ($\downarrow$) sweeps. Each of the networks consist of $N$ columns, each comprising an excitatory and an inhibitory population (Fig~\ref{fig:diagram}). 

\begin{figure*} 
          \centering
          \includegraphics[width=0.7\textwidth]{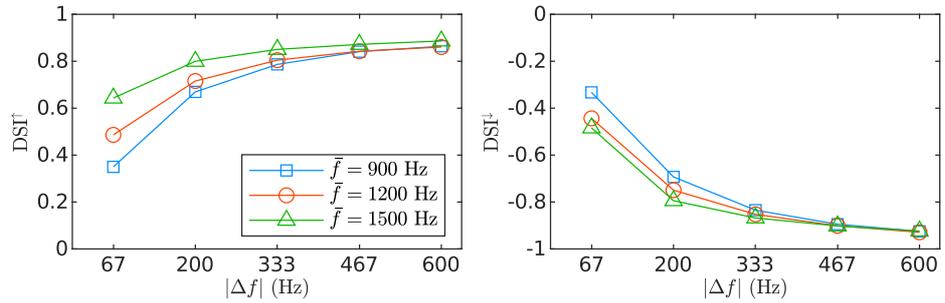}
          \caption{\textbf{Direction selectivity indices for the sweeps of the experiment.} DSI${\uparrow}$ and DSI$^{\downarrow}$ to sweeps with different $\bar{f}$ and $|\Delta f|$. DSI is defined as the proportional activity to up in comparison to down sweeps in a given network.}  
          \label{fig:dsiFixed}
        \end{figure*}

        To quantify direction selectivity, we used the standard direction selectivity index (DSI; e.g.,~\cite{Zhang2003}), defined as the proportion of the activity elicited in a network by an up sweep minus the activity elicited in the same network by a down sweep with the same duration and frequency span. An ideal network responding selectively to up sweeps will have a $\text{DSI} = +1$ and an ideal network responding selectively to down sweeps will have a $\text{DSI} = -1$. Similar DSI magnitudes are measured in the down and the up network (Fig~\ref{fig:dsiFixed}); systematically increasing DSI magnitudes were elicited by increasing $\bar{f}$ and $|\Delta f|$. Network selectivity to FM direction was robust across reparametrisations of the model, although deactivation of the feedback connections resulted in a 8.7($\pm1.5$)\% average decrease in DSI$^{\uparrow}$ and in a 9.7($\pm1.4$)\% average increase in DSI$^{\downarrow}$, indicating that the feedback connections sharpen direction selectivity.

      \subsubsection{Predictive mechanisms \label{sec:mod:model:nmda}}
        
        Once neurons in the sweep layer encode the sweep direction, feedback connections targeting the spectral layer facilitated the encoding of expected frequencies. Let $i$ be the population in the up-sweep network receiving inputs from a population in the spectral layer encoding a certain frequency $f_0$. Due to delayed excitation, the population $i$ becomes active when it detects an up sweep occurring in the neighbourhood of frequencies $f \leq f_0$. Although in some occasions the up sweep will culminate in $f_0$, in most of the cases $f_0$ will be only an intermediate step in the ascending succession of the sweep and thus the activation of $i$ would imply that populations in the spectral layer with best frequencies immediately higher than $f_0$ are likely to activate next. The top down mechanism of the model, encoded in the feedback projections stemming from the sweep layer and targeting the spectral layer, reduced the temporal constant of these populations using low-current feedback excitatory signals. Similarly, feedback connections stemming from a population $j$ in the down-network that received timely inputs from a spectral population with best frequency $f$ will target populations in the spectral network with best frequencies immediately lower than $f_0$. 

        NMDA receptors are typically responsible for conveying feedback excitatory information in the cerebral cortex \cite{Friston2001, Salin1995}; specifically, NMDA-deactivation results in a reduced feedback control in the auditory pathway \cite{Rauschecker1998}. Thus, while bottom-up drive was modelled using AMPA dynamics, feedback connections were modelled according to NMDA-like synaptic gating dynamics with a finite rising time constant \cite{Brunel2001}. Feedback current intensity was kept low in comparison to the bottom-up driver by enforcing NMDA conductivity to be much smaller than the AMPA conductivity (i.e., $J\NMDA \ll J\AMPA$).

\begin{figure}
          \includegraphics[width=\columnwidth]{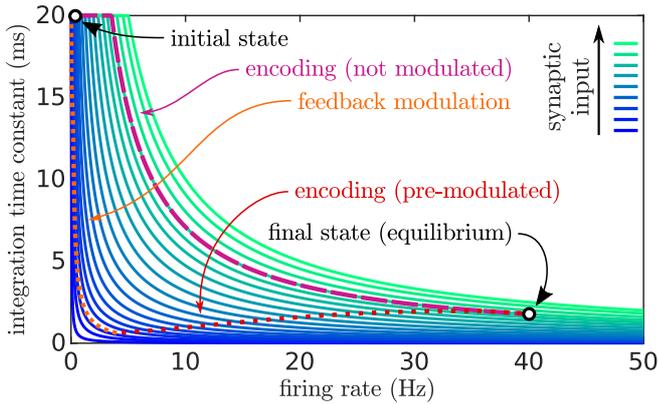}
          \caption{\textbf{Effect of the predictive feedback mechanism on the population time constants.} $\tau\supix{pop}(h, I)$ (green-blue solid lines) depends on the firing rate $h$ and the synaptic input $I$ (cf. Eq~\eqref{eq:taupop}). The figure portraits two different trajectories of the variable $\tau\supix{pop}(h, I)$ in the $(h, I)$ space, both starting at an initial state ($h \sim 0$ in the regime of spontaneous activity with no inputs) and finishing at a equilibrium state (with $h \sim 40$\,Hz). The dashed purple line shows a trajectory followed by the population when the forward synaptic input from the peripheral layer is plugged in without previous modulation. In this case, the population reacts slowly to the strong synaptic input, and eventually converges to equilibrium. The dotted lines (orange and red) show the trajectory of the same population in the presence of feedback modulation. The low-current modulatory inputs drive the population to a state with a low effective time constant without substantially increasing its firing rate (orange section of the trajectory). When the strong synaptic input from the auditory periphery is switched on (red section of the trajectory) the population reacts quickly to the synaptic, reaching equilibrium much faster than in the non-modulated case.}
          \label{fig:tauTrajectories}
        \end{figure}

        The low-current feedback signal modulates the population to elicit only a subtly higher firing rate than a not modulated population. Due to network effects captured in the mean-field model \cite{Ostojic2011}, this subtle activation driven by a low-current input resulted in a significantly lower effective integration time constant at the neuronal population level (Fig~\ref{fig:tauTrajectories}), causing the population to react faster to changes in the bottom-up input. This increased readiness reduced the metabolic cost of encoding expected frequencies and, since the population will spend more time in the high-firing-rate regime, it indirectly results in a stronger contribution of the frequencies expressed in the last part of the sweep to the probability distribution of pitch.

      \subsubsection{Reproduction of the sweep pitch shift \label{sec:mod:results:sweeps}}

\begin{figure*}
          \centering
          \includegraphics[width=\textwidth]{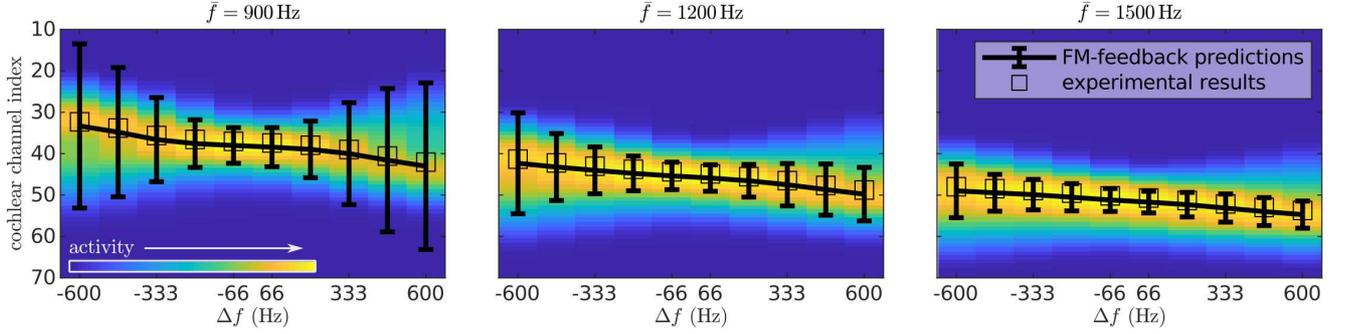}
          \caption{\textbf{Predictions of the FM-feedback spectral model for FM-sweeps.}  Heatmaps show the mean activation at different cochlear channels as simulated by a model of the auditory periphery in response to each sweep; units are arbitrary. Error bars point to the expected value and variance of the distribution across frequencies for each sweep. Each empty square denotes the expected channel elicited by a pure tone with the frequency of the average experimental data of the corresponding sweep.}
          \label{fig:swMod}
        \end{figure*}

        The FM-feedback spectral model explains $R^2 = 0.88$ of the variance of the experimental data (Fig~\ref{fig:swMod}). Moreover, there was a significant correlation between the variance of the model responses and the standard error of the experimental data ($r_p = 0.60$, $p = 0.0005$), indicating that the larger variability in the pitch shift observed for the larger $\Delta f$ can be understood as a consequence of a wider spread activation across the spectral populations.

\begin{figure*}
          \centering
          \includegraphics[width=0.7\textwidth]{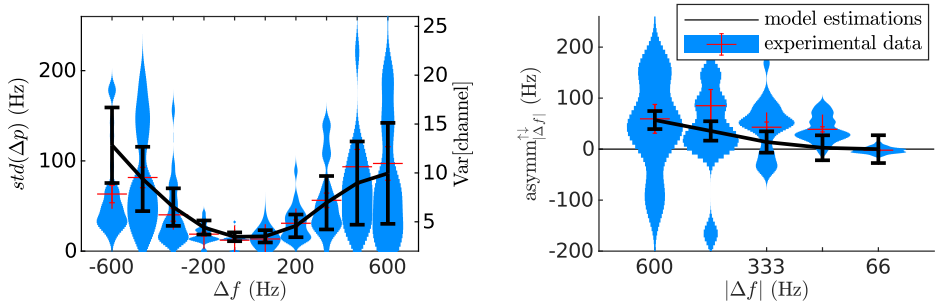}
          \caption{\textbf{Predictions of the FM-feedback spectral model for the up/down asymmetry.} Error bars show the model predictions of the up/down asymmetry coefficient asymm${\uparrow \downarrow}$ (see Eq~\eqref{eq:udAsymm}). Error bars are estimations of the standard error calculated based on the dispersion of the centroids for different $\bar{f}$ and the variance of the spectral distribution $\rho$ of each condition. Experimental data in the background is the same as in Fig~\ref{fig:swVar}, right.}
          \label{fig:swModAsymm}
        \end{figure*}

        Since up sweeps provoke a stronger overall activity in the auditory nerve \cite{Rupp2002}, facilitation currents were slightly higher for up than for down sweeps, resulting in a noticeable stronger absolute mean pitch shift for up than for down sweeps, reproducing the experimental data (Fig~\ref{fig:swModAsymm}). Note that this is not an obvious result of the model fitting for the data, as the expected difference between the absolute deviance $f_{\text{perceived}} - \bar{f}$ for up and down sweeps $E[\text{asymm}^{\uparrow \downarrow}] \simeq 24$\,Hz is significantly smaller than the average error of the model predictions with respect to the data ($E[\text{error}] \simeq 54$\,Hz).  

\begin{figure}
          \includegraphics[width=\columnwidth]{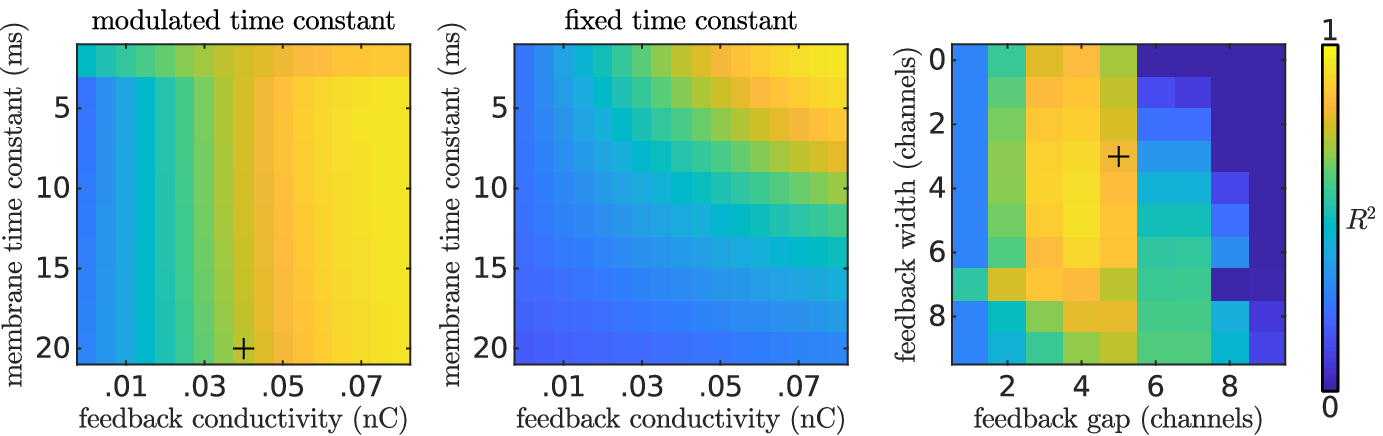}
          \caption{\textbf{Experimental fit in relation to the model parametrisation.} Shading matrices show the explained variance of the experimental data $R^2$ (bright yellow means $R^2 = 1$, dark blue means $R^2 = 0$) for different points in the parameter space. Unless stated otherwise, parameters not varied in the matrices correspond to the values listed in the Methods section. The two leftmost plots show the dependence of $R^2$ with the conductivity of the feedback connections and the dynamics of the excitatory population time constants. Different values of the nominal population's time constant were used to illustrate that the dynamic effect (rather than the resulting shorter time constant) is crucial to explain the experimental results; however, during the parameter tuning the temporal constant was constrained to $\tau\supix{memb} = 20$\,ms based on physiological observations \cite{McCormick1985}. The rightmost plot shows the dependence of $R^2$ on the width and reach ($w_{\omega s}$ and $\Delta_{\omega s}$, respectively; see Methods) of the feedback connections. Black crosses in the parameter space signal the final parametrisation.}
          \label{fig:pitchparspaceSw}
        \end{figure}

        To study the dependence of the fitness with the model's parameters we recomputed the explained variance $R^2$ across the parameter space of the model (Fig~\ref{fig:pitchparspaceSw}). The model explained the experimental data in a wide section of the parameter space, with an average $R^2$ across a 5-point diameter sphere around the final parameters of $E[R^2] = 0.78\pm0.03$. To show that fit of the model was not simply caused by an overall stronger activation provoked by the feedback currents, but by a decrease in the effective time constant of the populations, we also computed the dependence of $R^2$ with the conductivity of the feedback current $J\NMDA$ while keeping the population time constant $\tau$ fixed to $\tau = \tau^{\text{memb}}$ (see Methods). Even considering lower $\tau^{\text{memb}}$ than the physiologically valid nominal value $\tau^{\text{memb}} = 20\,$ms, without an adaptive $\tau$, much stronger NMDA currents ($J\NMDA \sim J\AMPA$) are necessary to drive the spectral distribution towards the experimental results.

      \subsubsection{Reproduction of previous experimental results \label{sec:mod:results:brady}}
        
\begin{figure}
          \includegraphics[width=\columnwidth]{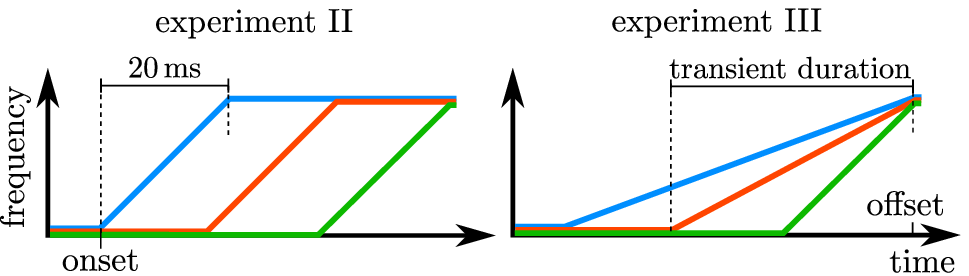}
          \caption{\textbf{Schematic view of the stimuli from \cite{Brady1961}.} In the stimuli from Brady's experiment II (left) the transient was fixed to a 20\,ms duration and its onset was systematically varied so that the transition falls at different segments of the stimulus. In the stimuli from Brady's experiment III (right) the stimulus offset was fixed at 90\,ms and the transient's onset varied between 10 and 50\,ms, resulting in transients of different durations. We extended the duration of these last stimuli to 95\,ms to prevent the ramping at the end of the stimulus from overlapping with the FM transient.}
          \label{fig:bradySchematic}
        \end{figure}

        We tested whether the FM-feedback spectral model was able to predict the pitch shift of additional data from the study by Brady and colleagues \cite{Brady1961}. We chose their stimuli because this was the only study that investigated the dependence of the pitch shift with properties different than $\Delta f$. Specifically, in the \emph{experiment II} from the original paper, Brady and colleagues considered FM-sweeps with a fixed 20\,ms transition between 1000\,Hz and 1500\,Hz that was located at six different positions within a 90\,ms stimulus (see schematics in Fig~\ref{fig:bradySchematic}, left). In the \emph{experiment III}, they used FM-sweeps in the same $\Delta f$ but with transitions of six different durations (see schematics in Fig~\ref{fig:bradySchematic}, right). All stimuli had the same duration (90\,ms) and frequency span (1000-1500\,Hz); in each of the two experiments there was a total of 12 stimuli (six up, six down).

\begin{figure*}
          \centering
          \includegraphics[width=\textwidth]{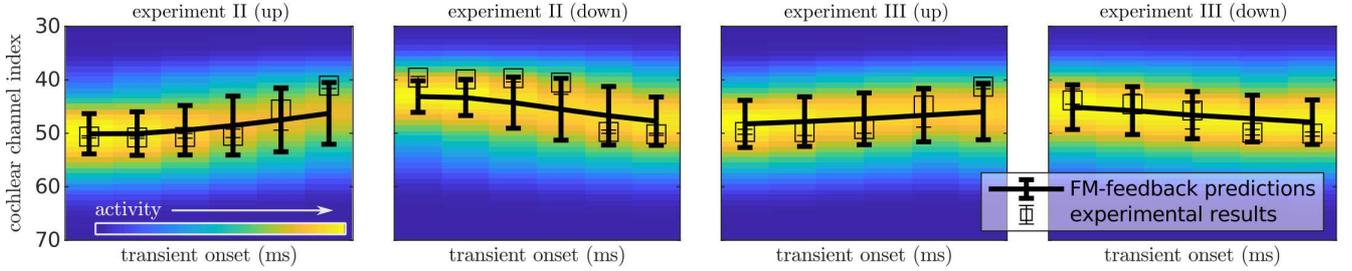}
          \caption{\textbf{Predictions of the FM-feedback spectral model for Brady's stimuli.} Shading matrices show the distribution of the activation across channels ($y$-axis) for different transient onsets ($x$-axis). Squares printed over the distributions mark the estimations of the experimental results in the channel space.}
          \label{fig:bradyPitch}
        \end{figure*}

        We compared the predictions of the FM-feedback spectral model with the experimental results reported in the original paper (Fig~\ref{fig:bradyPitch}). The experimental trend is well reproduced by the model. Predictions showed a strong Pearson's correlation with the reported sweep pitch shift across both experiments ($r_p = 0.87$, $p < 10^{-6}$) and a weaker but still significant correlation between the variance of the activation distribution $\rho$ and the experimental standard error ($r_p = 0.46, p = 0.03$). These correlations showed that the participant’s perception is well predicted by the model.

    \subsection{Testing the FM-feedback spectral model with a novel class of stimuli }


        The results described so far are in favour of the hypothesis that a feedback system between FM-sweep direction-encoding and frequency-encoding populations are responsible for the sweep pitch shift. To validate these findings, this section introduces a novel set of stimuli specifically designed to contest the main hypothesis of the model. The novel stimuli consist of concatenations of several single sweeps with the same properties as the stimuli used in the first experiment. We call them \emph{sweep train} in the following. Sweep trains present the same acoustical properties as the single sweeps used in the first behavioural experiment and should nominally elicit the same pitch percept as their single-sweep subcomponents. However, the FM-feedback spectral model predicts that the feedback system will only reduce the time constant of the spectral populations during the processing of the first sweep in the train, because they will already have an elevated firing rate (and thus a low effective time constant) during the processing of the subsequent sweeps in the train. Consequently, the model predicts that the sweep trains will elicit a much more subtle pitch shift than their single sweep counterparts. We tested this prediction in a perceptual experiment analogous to the first experiment.

      \subsection{Sweep trains show minimal sweep pitch shift}
      
\begin{figure*}
          \centering
          \includegraphics[width=\textwidth]{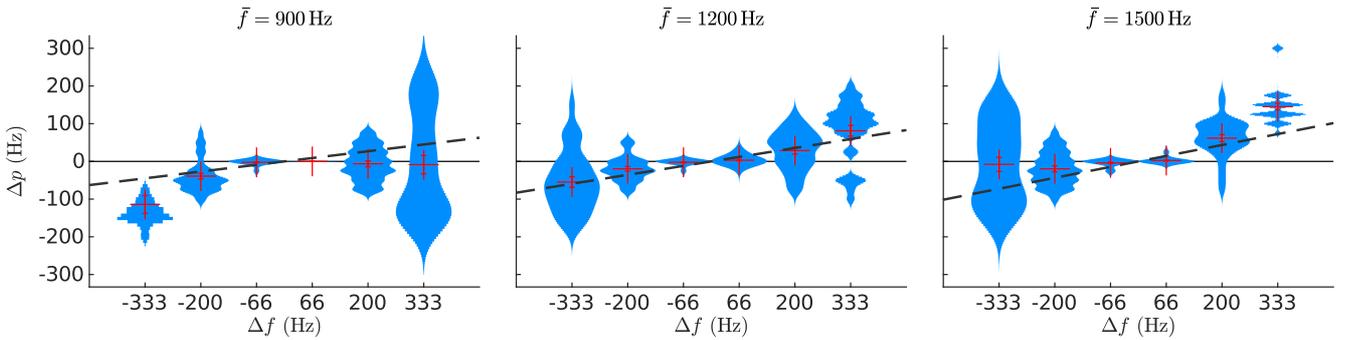}
          \caption{\textbf{Sweep pitch shift for sweep trains.} Kernel density estimations on the perceived pitch are plotted separately for each of the 18 sweep trains used in the experiment. The $y$-axis of each plot shows the magnitude of the sweep pitch shift $\Delta p$. The x-axis list the gaps of each of the sweeps. Red crosses show the mean and standard error of the data. Dark dashed lines show the group linear fit of the data.} 
          \label{fig:trPitch}
        \end{figure*}

        To ensure that each train was perceived as a single auditory object, we only used sweeps with $|\Delta f| \leq 333$\,Hz to ensemble the sweep trains, resulting in a total of $3\times6 = 18$ stimuli. The magnitude of the pitch shift depended $\Delta f$ (Fig~\ref{fig:trPitch}, Tab~\ref{tab:stats}, bottom). However, as qualitatively predicted by the FM-feedback spectral model, the effect sizes of the correlation were lower than in the single-sweep experiment (cf., Tab~\ref{tab:stats}, top). Data also showed much higher inter- and intra-subject variability than in the single-sweep experiment (Fig.~\ref{fig:slopes}). After completing the experiment, participants reported that the sweep train stimuli were harder to match than the single-sweep counterparts. Although trains with small $\Delta f$ were generally perceived as continuous tones, subjects reported that a few trains (putatively those with the largest $\Delta f$) elicited a ringing-phone-like percept. Stimuli are available in the supporting information (\nameref{sounds}).

        Sweep-train stimuli show only a subtle up/down asymmetry that did not reach statistical significance ($p = 0.67, p = 0.96, p = 0.36$ for $|\Delta f| = 333$, $|\Delta f| = 200$, $|\Delta f| = 66$, respectively; according to two-sided Wilcoxon signed rank tests with 24 samples per condition).

      \subsection{The FM-feedback spectral model explains the diminished pitch shift in the sweep trains}

\begin{figure*}
          \includegraphics[width=\textwidth]{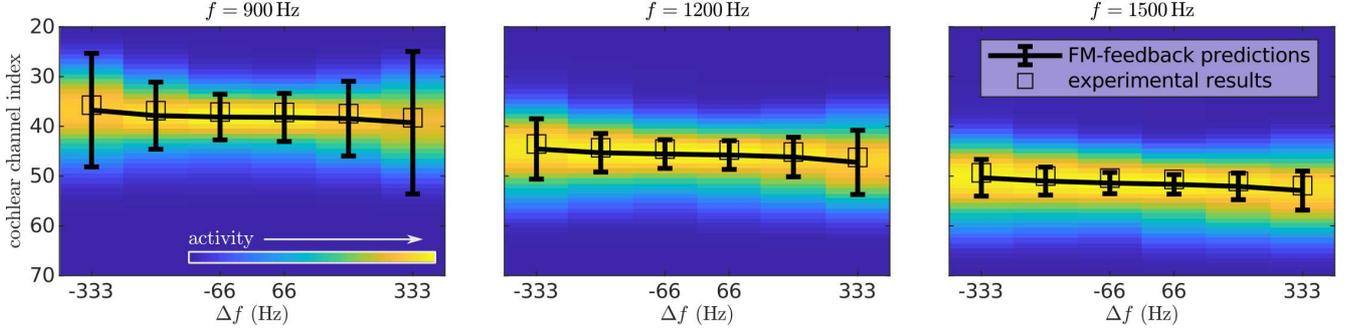}
          \caption{\textbf{Predictions of the FM-feedback spectral model for sweep trains.} Shading matrices show the distribution of the activation across channels ($y$-axis) for different sweep $\Delta f$ ($x$-axis). Squares printed over the distributions mark the expected channel $E[k]$ as defined in Eq~\eqref{eq:spectralPitch}. Solid error bars are estimations of the experimental results in the channel space. The expected value agrees with the experimental data. Moreover, stimuli with larger $\Delta f$ seem to elicit wider activation distributions than stimuli with smaller $\Delta f$, mirroring the generally larger variance observed in the behavioural data corresponding to the larger $\Delta f$.}
          \label{fig:trMod}
        \end{figure*}

        Next, we assessed the ability of the FM-feedback spectral model to quantitatively explain the effect size of the pitch shift observed in the sweep trains. The fit with the experimental data was comparable to that of the single sweep stimuli: the model explained $R^2 = 0.83$ of the variance of the data (Fig.~\ref{fig:trMod}) and the response distribution's expected value was strongly correlated with the observed pitch shift ($r_p = 0.99, p < 10^{-18}$;).

        As in the first experiment, the variance of the experimental data was strongly correlated to the width of the model responses ($r_p = 0.60, p = 0.0005$; Fig~\ref{fig:swModAsymm}, left).

        \paragraph{}

        \begin{figure*}
          \centering
          \includegraphics[width=0.7\textwidth]{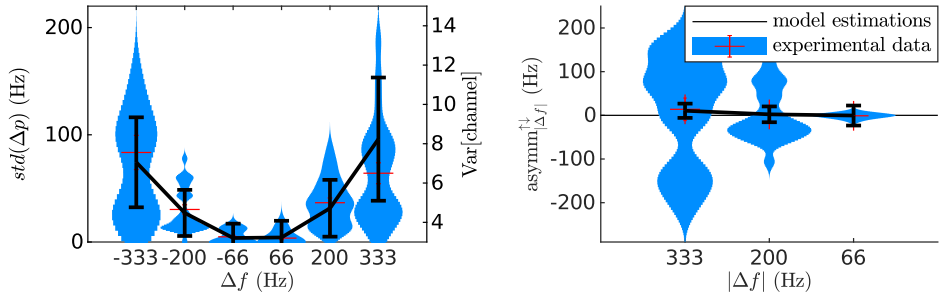}
          \caption{\textbf{Predictions of the FM-feedback spectral model for the variance and up/down asymmetry in sweep trains.} Left: Kernel density estimations of the intra-subject standard deviation of the sweep pitch shift $\Delta p$, plotted separately for the different frequency spans $\Delta f$. Each sample in the distributions corresponds to the standard deviation of the perceived pitch of a sweep in one subject (i.e., in each distribution there are $8 \times 3 $ points, one for each subject and $\bar{f}$). The variance is monotonically correlated to the absolute gap $|\Delta f|$ ($r_s = 0.63$, $p < 10^{-27}$). Right: Kernel density estimations of the up/down asymmetry distributions as defined in Eq~\eqref{eq:udAsymm}. Each sample of the distributions corresponds to the difference of the average absolute deviation from centre frequency between up and down sweeps of the same $|\Delta f|$ for a given subject and centre frequency ($N = 8\times3 = 24$). Red crosses show the mean and the standard error of the data. Error bars show the model predictions of the up/down asymmetry coefficient asymm${\uparrow \downarrow}$ (see Eq~\eqref{eq:udAsymm}). Error bars are estimations of the standard error calculated based on the dispersion of the centroids for different $\bar{f}$ and the variance of the spectral distribution $\rho$ of each condition. Cf.,~Fig~\ref{fig:swModAsymm}}
          \label{fig:trModAsymm}
        \end{figure*}

        Last, we tested whether the different up/down asymmetry (asymm$^{\uparrow \downarrow}$) observed in the single sweeps and sweep train data could be quantitatively explained by the FM-feedback spectral model. In the single-sweep data, the model predicts a stronger pitch shift magnitude $|\Delta p|$ for up sweeps (Fig~\ref{fig:swModAsymm}) because, due to the compensation for the delay introduced by the basilar membrane in response to low frequencies, these elicit a more synchronous and stronger peak activation in the auditory nerve \cite{Rupp2002}, resulting in larger feedback currents. Qualitatively, a much weaker asymmetry was expected in the sweep-train data, since the effects of the feedback system are virtually absent during the processing of the ending four fifths of the stimuli. Modelling results on the up/down asymmetry closely reproduced the empirical data (Fig~\ref{fig:swModAsymm}), fully explaining the observed differences between the two families of stimuli.

  \section{Discussion}


      In this work we have built a model describing how cross-feature feedback between two different representations of frequency modulation gives rise to a puzzling perceptual effect. This contrasts with the classical view of FM encoding as a bottom-up process \cite{Skorheim2014}. The predictive feedback proposed in this work aids efficient encoding of frequency modulation by decreasing its metabolic cost \cite{Alexandre2018}, shortening its processing time \cite{Jaramillo2011, Mazzucato2019}, and enhancing direction selectivity.

    \subsection{Bottom-up pitch models and pitch codes}
      
      Two codes of pitch-related information are available in the auditory nerve at early stages of the auditory pathway: 1) the spectral code, produced by the spectral decomposition of the stimuli performed by the basilar membrane; and 2) the temporal code, comprised in the spike timings of the neurons across the auditory nerve that are phase locked to the stimulus waveform (see~\cite{Oxenham2013} for a review).

      Our simulations showed that current modelling approaches based on the temporal code do not suffice to explain the pitch of the FM-sweeps used in the experiments. This is most likely a consequence of the fast change rate in the periodicities of fast FM stimuli. Typically, pitch decisions based on the auditory nerve temporal code are made after integrating over four cycles of the period of the stimuli \cite{Tabas2019, Wiegrebe2001}, coinciding with the duration threshold for accurate pitch discrimination \cite{Krumbholz2003}. However, our stimuli presented an average change of $\sim25$\,Hz across four repetitions of their average frequency, making this integration virtually impossible. Thus, the FM-feedback spectral model assumes that the pitch of FM sweeps and pure tones is encoded in a spectral representation, siding with the idea that spatial information can still play a crucial role in pitch processing. 

      The bottom-up integration of the spectral representation, cornerstone of the classical spectral theories of pitch \cite{Helmholtz1863}, predicted a sweep pitch shift in the opposite direction of the experimental data; i.e., a shift towards the frequencies expressed at the beginning of the sweep. This is a direct consequence of the global adaptation effects experienced in the auditory nerve after the first few milliseconds of the stimuli \cite{Zilany2009}. Even without such adaptation, the plain integration proposed by the spectral models would predict a null sweep pitch shift. Feedback modulation facilitating the encoding of the predictable parts of the sweeps is thus crucial to account for the experimental data.

    \subsection{Relation to predictive coding and hierarchical processing strategies}

      The presence of predictive feedback modulation in the subcortical sensory pathway has been shown before in humans \cite{Suga2012, Tabas2020} and non-human mammals \cite{Malmierca2015}. Previous studies often interpreted it in the context of the predictive coding framework \cite{Mumford1992, Rao1999, Friston2005}, a theory of sensory processing that postulates that sensory information is encoded as prediction error; i.e., that neural activity at a given level of the processing hierarchy encodes the residuals of the sensory input with respect to a generative model encoded higher in the hierarchy. 

      The FM-feedback spectral model can also be understood in the light of this formalism: it presents three hierarchical layers of abstraction (the inputs from the peripheral system, the frequency network, and the sweep network), and the two top layers perform predictions on the sensory input incoming at the immediately lower representation of the hierarchy. In the case of the frequency network, the temporal integration can be interpreted as the prediction that the input's distribution across cochlear channels will change with a much longer time constants as that of the fluctuations introduced by neuronal noise. However, unlike the classical predictive coding microcircuit where predictions and prediction error are kept in separate neural ensembles \cite{Bastos2012}, the frequency and sweep network simultaneously hold a representation that is both, descriptive for their own representation and predictive for the immediately lower representation of the hierarchy. 

      Combining predictions and representations in the same neural code solves some of the open questions of classical predictive coding architectures recently summarised by Denham and Winkler \cite{Denham2018}: i) ``what precisely is meant by prediction?'', ii) ``which generative models [within the hierarchy] make the predictions?'', and iii) ``what within the predictive framework is proposed to correlate with perceptual experience?''. In the FM-feedback spectral model, the predictions can be summarised as the probability distribution of patterns of activation expected to come next in the lower level given what has been encoded so far in the higher level. These conditional probability distributions are hardcoded in the top-down connections stemming from the neurons holding the high-level representation and targeting the neurons holding the lower level representations. Such connectivity patterns would represent the statistics between the representations in the two levels if they were naturally formed through synaptic plasticity after sufficient exposure to the stimuli. Last, the perceptual experience in the FM-feedback spectral model is encoded in the activation along the two hierarchical stages, which encode different aspects of the stimuli.

      Another key difference between the FM-feedback spectral model's architecture and the classical predictive coding microcircuit is that, rather than encoding the residuals of the spectral representation with respect to the FM-sweep representation, neurons in the spectral layer simply encode the spectral content of the stimulus. However, since the decoding of the predictable parts of the stimuli is faster and its metabolic cost lower, predictability potentially ensues a significant decrease on the amount of signal produced during the encoding. Such mechanism would explain why even expected stimuli, for which the residual should theoretically be zero, do still evoke measurable responses (as in, for instance, stimulus-specific adaptation \cite{Ulanovsky2003, Malmierca2015}).

    \subsection{Comparison with previous measurements of the sweep pitch shift}

      Our experimental findings qualitatively replicated the sweep pitch shift effect found in previous studies; namely, we found that the pitch elicited by FM-sweeps was biased towards the frequencies spanned in the ending part of the sweeps \cite{Brady1961}, and that the perceptual bias is monotonically related to the frequency span $\Delta f$ \cite{Nabelek1970, Rossi1978}. On average, we estimated a putative linear relation between the pitch shift $\Delta p$ and $\Delta f$ of around $m \simeq 0.38$, slightly higher than Brady's \cite{Brady1961} ($m \simeq 0.34$ with transitions of 50\,ms) and Nabelek's \cite{Nabelek1970} ($m \simeq 0.32$ with transitions of 40\,ms) reports, and significantly higher than Rossi's \cite{Rossi1978} ($m = 1/6 \simeq 0.17$ with transitions of 200\,ms) estimation. Since Rossi's transitions were 5 times longer than ours, the estimations are difficult to compare. However, the disagreement seems to indicate that the pitch shift is stronger with shorter durations. This observation would be fully compatible with the mechanism of predictive facilitation described in the FM-feedback spectral model: since the time to decode FM direction is independent of sweep duration, whilst only the most posterior part of the stimulus is facilitated in the short sweeps, in a long sweep the facilitation currents would affect a much larger portion of the sound, potentially including frequencies occurring before $\bar{f}$. 

      The subtle disagreement of our predictions with Brady's and Nabelek data has three possible explanations: 1) the differences are a result of the studies having relatively low sample sizes in comparison with the high inter-subject variability of the effect (see~\ref{fig:slopes}); 2) Brady's and Nabelek's studies do not report any participant selection criteria: perhaps the inclusion of listeners that were unable to perform the match resulted in experimental results biased towards a null effect (i.e., towards $\Delta p = 0$); and 3) Nabelek and Brady used analogue synthesisers to produce their stimuli, resulting in sweeps with a richer spectral contour than our digital FM-sinusoids, which might have resulted in a weaker effect (Fig~6 in \cite{Brady1961} indicates that the spectral properties of the sweep do indeed affect the pitch shift: sweeps of the same duration, spectral scope and $\Delta f$ produced different sweep pitch shift magnitudes).

      \paragraph{}
      
      The FM-feedback spectral model also provides for a mechanistic interpretation of these previous results. In Brady's experiment II, the transient duration of the sweep is kept constant but its onset is varied across the stimulus duration. When the transient is located near the beginning of the stimulus, the greatest part of the sounds excites neurons encoding frequencies close to the posterior parts of the transient pushing the distribution of the responses towards the ending frequencies of the sweep $f_1$. This shift is larger than it would be expected for a sound without a transient because of the feedback modulation of the later frequencies exerted by the sweep network. When the transient is located at the very end of the stimulus, the longer portion of the stimulus exciting $f_0$ compensates for the shift in the frequency distribution, bringing the perceived pitch closer to the starting frequencies of the stimulus.

      In Brady's experiment III, the transient's onset is kept constant and it is the duration of the transient that is varied. The decreased sweep pitch shift observed for shorter in comparison to longer transient durations can be explained by the FM-feedback spectral model as a consequence of the stimuli presenting a larger segment with the initial frequency, thus shifting the distribution of the responses towards $f_0$.

    \subsection{FM encoding and physiological location of the sweep and spectral layers}

      FM direction selectivity was modelled according to the principles of delayed excitation \cite{Razak2008, Ye2010, Kuo2012}. Although both delayed excitation and sideband inhibition contribute to direction selectivity in the mammalian auditory pathway \cite{Williams2012, Fuzessery2011, Geis2013}, neural modelling based on each of the two mechanisms produces similar representations \cite{Skorheim2014}. We chose to use delayed excitation alone in order to prevent a disproportionate inflation in the number of free parameters of the model.

      Although we did not attempt to model FM rate selectivity, the FM-feedback spectral model's DSIs monotonically increased with $\Delta f$, a property that could be exploited in further developments of the model to encode modulation rate. FM rate encoding has been reported in mice \cite{Geis2013, Trujillo2013}, rats \cite{Lui2003} and more extensively in bats (e.g.,~\cite{Gittelman2011, Fuzessery2011}). 

      The earliest neural centre within the auditory pathway showing FM direction selectivity in mammals is the inferior colliculus \cite{Kuo2012, Geis2013, Li2010, Hage2003}, although subsequent nuclei (medial geniculate body \cite{Kuo2012, Lui2003} and auditory cortex \cite{Issa2016, Trujillo2013, Ye2010, Li2010, Zhang2003}) show generally stronger DSIs. Thus, the sweep layer postulated in the FM-feedback spectral model could be implemented even at early stages of the auditory hierarchy. Similarly, since all the nodes in the ascending auditory pathway contain tonotopically arranged nuclei, the spectral layer could be putatively located as early as in the cochlear nucleus. The physiological location of the mechanisms described here remains an open question.

  \section{Conclusion}

    In this work we have harnessed a well-established perceptual phenomenon to inform a model of FM direction encoding. We have shown that neither phenomenological nor mechanistic bottom-up models of auditory processing are able to explain the experimental data. We concluded that FM direction-selective neurons must alter the way that spectral information is encoded via a feedback mechanism. The main contribution of this work is a specific theory of how this feedback modulation might be exerted. Given the paramount role played by fast FM-sweeps in speech, the predictive mechanisms described here could be part of a larger hierarchical network responsible for the encoding of speech sounds in the human auditory pathway.

  \section{Materials and methods}

    \subsection{Measuring the sweep pitch shift in single sweeps}

      \subsubsection{Participants \label{sec:sw:methods:listeners}}

        8 participants (4 female), aged 22 to 31 (average 26.9) years old, were included in the study. They all had normal hearing thresholds between 250\,Hz and 8\,kHz ($< 25\,$dB HL) according to pure tone audiometry (Micromate 304, Madsen Electronics). All reported at least five years of musical experience, but none of them was a professional musician. 

        The 8 participants were derived from a larger set of 22 candidates. Candidates were screened by a first behavioural test assessing their capacity to match pure tones against pure tones, and then by a second test measuring their consistency when matching sweeps against pure tones (see details bellow). From the 14 excluded participants, one failed the first test and 13 failed the second test. 6 of the excluded participants reported no previous musical experience; the remaining 8 had at least five years of musical training. 

      \subsubsection{Stimuli \label{sec:sw:methods:stimuli}}

        Stimuli were $50$\,ms long frequency-modulated sweeps. Frequency was kept constant during the first and final 5\,ms of the sweeps. The modulation was asymptotic and carried out in 40\,ms. Stimuli were ramped-in and damped-out with 5\,ms Hanning windows overlapping the sections with constant frequency.

        There were 30 single sweeps with 10 linearly distributed frequency spans $\Delta f \in [-600, 600]$\,Hz and 3 average frequencies $\bar{f} \in \{900, 1200, 1500\}$\,Hz. For each sweep with a given $\Delta f$ and $\bar{f}$, the initial and final frequencies were $f_0 = \bar{f} - \Delta f / 2$ and $f_1 = \bar{f} + \Delta f / 2$.

      \subsubsection{Experimental design \label{sec:sw:methods:design}}

        Each trial consisted of a sequential presentation of a target sweep and a probe pure tone. After the presentation, the participant was asked whether the second sound evokes a higher, equal, or lower pitch percept than the first sound. Participants were allowed to replay the sounds as many times as needed in case of doubt. After the response, the software adjusted the frequency of the probe tone by increments of $\pm \epsilon = \pm 25$\,Hz, bringing the pitch of the sound closer to the participant’s percept (e.g., if the participant judged the target sweep as having a lower pitch than the probe tone, the frequency of the probe tone was reduced by 25\,Hz). This procedure was repeated until the participant reported that the two sounds evoked the same pitch percept. Then, the frequency of the matched pure tone was stored as the perceived pitch of the sweep reported in that trial, and a new trial with a new target sweep began. The initial frequency of the probe tone was sampled from a Gaussian distribution centred on the average frequency $\bar{f}$ of the target sweep. 
        
        Each of the 30 sweeps was matched four times, so that there were 120 trials in total in the experiment. The relative order of the probe tone and the target sweep was reversed in half of the trials to assess if presentation order affects the sweep pitch shift. Thus, the experiment can be described as a 10 (10 different frequency spans) $\times$ 3 (3 average frequencies) $\times$ 2 (probe played first or last) factorial design.

      \subsubsection{Experimental procedure \label{sec:sw:methods:structure}}

        Before the experiments, all potential participants performed a brief training to ensure that they had understood the task. The trial structure of the training was exactly the same as in the experiment, but both probe and target consisted of pure tones. During the training, the software provided feedback after each trial informing the participant whether the response was correct or incorrect. The training was divided in batches of six trials, and it concluded when the participant correctly matched the pitch of every trial in one batch. Most participants completed the training in the first batch. 

        After the training, we evaluated the consistency of each potential participant when matching the pitch of FM-sweeps. During the evaluation, participants undertook a block of 12 trials consisting in 4 repetitions of the same 3 sweeps: $\{\Delta f = 67\,\text{Hz}, \bar{f} = 900\,\text{Hz}\}$, $\{\Delta f = -200\,\text{Hz}, \bar{f} = 1200\,\text{Hz}\}$, and $\{\Delta f = -67\,\text{Hz}, \bar{f} = 1500\,\text{Hz}\}$. We chose these sweeps to ensure consistency across several $\bar{f}$ and $\Delta f$ while keeping $|\Delta f|$ small enough to ensure that the sweeps would elicit an unequivocal pitch percept \cite{Hart1990}. After the completion of this block, we scored the participant's pitch matching consistency as the inverse of the average of the absolute differences between the reported pitch in each sweep. Only participants with an average deviation smaller than twice the frequency increment step $2\epsilon = 50$\,Hz were included in the experiment. We assumed that participants with larger deviations were either unable or unwilling to perform consistent pitch judgements on sweeps, and thus their inclusion would contaminate the data with random guesses that could bias our estimations of the sweep pitch shift towards $\Delta p = 0$.

        The 8 included participants matched the remaining 27 sweeps in four additional blocks. No sweep type was repeated within a single block, and all sweeps were presented 4 times across the entire experiment, resulting in 27 trials per block. The order of the sweeps within each block was randomised and the relative position of the probe tone with respect to the target stimulus was pseudorandomised so that half of the trials in each block were presented in each direction. Participants were instructed to take rests between blocks and were allowed to take as many shorter rests between trials as needed. To encourage precision, a 5\euro{} award was offered to participants that kept their self-consistency along the main experiment with the same criterium as in the evaluation: a smaller variance than $2\epsilon = 50$\,Hz within each sweep type. Only sweeps expected to yield the most unequivocal pitch sensation according to Hart's law \cite{Hart1990} (i.e., $|\Delta f| \leq 200$\,Hz) were used to compute the overall self-consistency; participants were however unaware of this. Participants typically completed the experiment within 3 hours.

    \subsection{Measuring the sweep pitch shift in sweep trains}

      \subsubsection{Participants \label{sec:tr:methods:listeners}}

         The same 8 participants who completed the first experiment were invited to repeat the measurements with the new stimuli.

      \subsubsection{Stimuli  \label{sec:tr:methods:stimuli}}

\begin{figure*}
          \includegraphics[width=\textwidth]{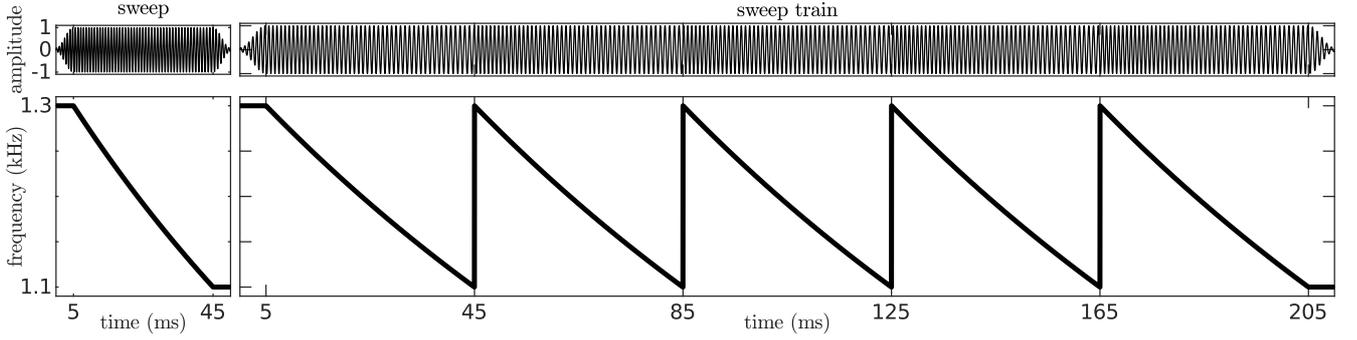}
          \caption{\textbf{Examples of the stimuli.} Waveform $s(t)$ and instantaneous frequency $f(t)$ of the sweep with $\bar{f} = 1200$\,Hz and $\Delta f = -200$\,Hz and its corresponding sweep train.}
          \label{fig:stim}
        \end{figure*}

        Stimuli were concatenations of 5 sweeps adding up to a total of $250$\,ms (sweep trains; see Fig~\ref{fig:stim}). The sweeps were taken from a subset of 18 elements from the first experiment with 6 different frequency spans $\Delta f \in [-333, 333]$\,Hz. To ensure continuity of the stimulus waveform, the sweeps were concatenated in the frequency domain. 5\,ms Hanning windows were applied only at the very beginning and very end of the sweep trains.

      \subsubsection{Experimental design \label{sec:tr:methods:design}}

        The matching procedure was the same as in the first experiment: the participants matched the pitch of the sweep trains to probe pure tones whose frequency they could adjust with the aid of a computer software. To ensure that there were no effects of stimulus duration, the probe tones had the same duration as the sweep trains (i.e., 250\,ms). As in the first experiment, each of the 18 sweep trains was matched four times, so that there were 72 trials in the second experiment. The relative order of the probe tone and the target sweep train was also reversed in half of the trials. Thus, the second experiment can be described as a 6 (different frequency spans) $\times$ 3 (average frequencies) $\times$ 2 (probe played first or last) factorial design.

      \subsubsection{Experimental procedure \label{sec:tr:methods:structure}}
          
        Since the participants were already familiar with the task, the experiment contained no training. Four repetitions of the $18$ sweep-trains were distributed across $5$ blocks following the same principles as described in the first experiment. Participants typically completed the second experiment within 2 hours.

    \subsection{Bottom-up models of pitch}

      \subsubsection{Spectral models of pitch processing \label{sec:sw:models:spectral}}

        The responses at the auditory nerve were computed with a model of the peripheral auditory system \cite{Zilany2014, Zilany2009}. The model's output represents the expected firing rate $p_n(t)$ in a fibre of the auditory nerve associated with the $n$th cochlear channel ($n = 1, 2, \dots, N$) at an instant $t$. The frequency range of the cochlear model was discretised in $N = 100$ channels, spanning frequencies from $f_{\text{min}} = 125$\,Hz to $f_{\text{max}} = 10$\,kHz.

        The perceived pitch corresponded to the expected cochlear channel $k$, $E[k]$, according to a probability distribution $\rho$ derived from the integral of $p_n(t)$ over the duration of the stimulus $L$:

        \begin{equation} \label{eq:peripheralPitch}
          E[k] = \sum_n n \rho_n \quad \text{with} \quad 
            \rho_n = \frac{\int_0^{L} dt \, p_n(t)}{\sum_n \int_0^{L} dt \, p_n(t)}
        \end{equation}

        To compare the predictions of the model with the experimental data, we also computed the expected channels $E[k]$ associated to pure tones with the frequency of the average perceived pitch of each sweep. 

      \subsubsection{Temporal models of pitch processing \label{sec:sw:models:temporal}} 

        The SACF used in this work follows the original formulation by Meddis and O'Mard \cite{Meddis1997, Meddis2006}. Essentially, this model poses the existence of an array of $M$ periodicity detectors responding more saliently to a preferred period $\delta t_m$. The instantaneous firing rate $A_m(t)$ of the $m$th periodicity detector ($m = 1, 2, \dots, M$) follows:

        \begin{equation} \label{eq:sacf}
          \tau\supix{SACF}_m \dot{A}_m(t) = - A_m(t) + \sum_n p_n(t) \, p_n(t - \delta t_m)
        \end{equation}

        \noindent where the auditory nerve activity $p_n(t)$ in the cochlear channel $n$ at an instant $t$ is computed as in the previous section. The characteristic periods $\delta t_m$ are uniformly distributed between $\delta t_m = 0.5$\,ms and $\delta t_m = 30$\,ms, which allows the model to capture periodicities corresponding to frequencies between 2\,kHz and 135\,Hz up to four lower harmonics. We kept a fixed integration constant $\tau\supix{SACF}_m = 2.5\,$ms; using variable $\tau\supix{SACF}_m$ that depend linearly on $\delta t_m$ (see details in~\cite{Balaguer2008, Wiegrebe2004}) did not result in substantial changes in our results.

        Stimuli presenting periodicities at a certain frequency $f$ typically elicit peaks of activation in the detectors tuned to the preferred period $\delta t_m = 1 / f = T_0$ and to the periods corresponding to all subsequent lower harmonics $\delta t_m = 2\,T_0 = T_1$, $\delta t_m = 3\,T_0 = T_2$, etc. Thus, evidence for the period $T$ at an instant $t$, $B(t)_T$ can be represented as the $B(t)_T = \sum_{m \in\{\mathcal M_T\}} A_m(t)$, where $\mathcal{M_T}$ are the indices of the periodicity detectors tuned to $T$, $2T$, $3T$, etc. (i.e., $\mathcal{M_T} = \{m: \delta t_m = n T \quad \forall \, n \in [1, 2, 3, ...]\}$). We estimated $B(t)_T$ using four harmonics; extending or reducing the number of harmonics used to estimate $B(t)_T$ did not significantly alter our results.

        The perceived pitch corresponded to the expected period $T$, $E[T]$, according to a probability distribution $\rho$ derived from the integral of $B_T(t)$ over the duration of the stimulus $L$:

        \begin{equation} \label{eq:temporalPitch}
          E[T] = \sum_T T \rho_T \quad \text{with} \quad 
            \rho_T = \frac{\int_0^{L} dt \, B_T(t)}{\sum_n \int_0^{L} dt \, B_T(t)}
        \end{equation}

        To compare the predictions of the model with the experimental data, we computed the expected period $E[T]$ associated to pure tones with the frequency of the average perceived pitch of each sweep.

    \subsection{Details on the predictive model of FM encoding}
    
      \subsubsection{Spectral layer and pitch estimations}

        The spectral layer consists on an array of $N = 100$ neural populations that integrate the output of the peripheral model. Neural populations are modelled according to a mean-field derivation \cite{Wong2006} that, although it was first formulated to describe dynamics in cortical regions dedicated to visual decision making, have been successfully used to describe the dynamics of many different cortical areas (e.g.,~\cite{Deco2013}). The firing rate $h_n(t)$ of the $n$th ensemble follows the dynamics of a leaky integrator:

        \begin{equation} \label{eq:firingRate}
          \tau\supix{pop} \, \dot{h}^f_n(t) = -h^f_n(t) + \phi(I^f_n(t))
        \end{equation}

        \noindent where $\tau\supix{pop}$ are adaptive time constant:

        \begin{equation}
          \tau\supix{pop}_{e,i}(h, I) = \tau\supix{memb}_{e,i} \, \Delta_T 
                                 \frac{\partial_{x}\phi(x)\arrowvert_{x = I}}{h}
          \label{eq:taupop}
        \end{equation}

        \noindent $\Delta_T = 1\,$mV is the size of the spike initialisation of the neural model and $\tau\supix{memb}_e = 20\,$ms and $\tau\supix{memb}_i = 10$\,ms \cite{McCormick1985} are the neural membrane time constants for excitatory and inhibitory populations, respectively. Using adaptive integration time constants makes the populations to react faster to changes when they are marginally active and have weak synaptic inputs, a behaviour often reported in tightly connected populations of neurons \cite{Ostojic2011}. This component is the key of the feedback mechanism used to increase the responsiveness of the populations encoding the expected parts of the sweeps (Fig~\ref{fig:tauTrajectories}). The analytic formulation of $\tau\supix{pop}(h, I)$ stems from a theoretical study of networks of exponential-integrate-and-fire neurons \cite{Ostojic2011}.

        Inputs $I^f_n(t)$ were modelled with AMPA synaptic dynamics \cite{Brunel2001}. AMPA synapses present short time constants that are able to preserve the fine temporal structure of auditory input, and thus are the major receptor type conveying bottom-up communication in the auditory pathway (e.g.,~\cite{Golding2012}). 

        \begin{equation} \label{eq:fInputBottomUp}
          I^f_n(t) = J\subix{in}\AMPA \sum_k \omega\supix{in}_{nk} S\subix[,k]{in}\AMPA{}(t)
        \end{equation}

        \noindent We allowed some dispersion in the propagation from the peripheral model to the spectral layer by using a Gaussian-shaped connectivity matrix:

        \begin{equation}
          \omega\supix{in}_{nm} = \frac{1}{\sqrt{\sigma\subix{in}}} 
                                  e^{-\frac{(m - n)^2}{2 \sigma\supix{in}}}
        \end{equation}

        \noindent where the normalisation factor $\sqrt{\sigma\subix{in}}$ ensures that the total input to a population under a uniform peripheral input remains the same regardless of the dispersion $\sigma\subix{in}$. The synaptic gating variable $S\subix[,n]{in}\AMPA{}(t)$ follows \cite{Brunel2001}:

        \begin{equation}
          \tau\AMPA{} \dot{S}\subix[,n]{in}\AMPA{}(t) = 
                -S\subix[,n]{in}\AMPA{}(t) + p_n(t)
        \end{equation}

        \noindent Note that we used the index $^f$ to denote variables in the spectral layer. The perceived pitch corresponded to the expected cochlear channel $k$, $E[k]$, according to a probability distribution $\rho$ derived from the integral of $p_n(t)$ over the duration of the stimulus $L$ (cf. Eq~\eqref{eq:peripheralPitch}):

        \begin{equation} \label{eq:spectralPitch}
          E[k] = \sum_n n \rho_n \quad \text{with} \quad 
            \rho_n = \frac{\int_0^{L} dt \, h^f_n(t)}{\sum_n \int_0^{L} dt \, h^f_n(t)}
        \end{equation}

        The time constant $\tau\AMPA{} = 2$\,ms was taken from the literature \cite{Brunel2001}. The effective conductivity $J\subix{in}\AMPA{} = 0.38\,$nA was manually tuned within the realistic range such that the peripheral system would elicit firing rates on the range $5\,\text{Hz} \geq h_n(t) \geq 100$\,Hz in the integrator ensembles. The transfer function $\phi(x) = (c x - I_0)/(1 - e^{-g (c x - I_0)})$ and its parameters, empirically derived for networks of integrate-and-fire neurons, were taken from~\cite{Wong2006}.      
    
      \subsubsection{Sweep layer and direction selectivity}

        The sweep layer consists on four arrays of $N = 100$ neural populations following the same dynamics described in the previous section (i.e., Eq~\eqref{eq:firingRate}). From the four arrays, two (one excitatory, one inhibitory) are tuned to up sweeps, and two (again, one excitatory and one inhibitory) are tuned to down sweeps (Figure~\ref{fig:diagram}). The instantaneous firing rate of the \emph{up} ($h^{\uparrow e}_n(t), h^{\uparrow i}_n(t)$) and \emph{down} ($h^{\downarrow e}_n(t), h^{\downarrow i}_n(t)$) neural population , with \emph{up} ($I^{\uparrow e}_n(t), I^{\uparrow i}_n(t)$) and \emph{down} ($I^{\downarrow e}_n(t), I^{\downarrow i}_n(t)$) synaptic inputs, respectively. Although the transfer functions $\phi(x)$ are the same for all the ensembles, the parameters $c$, $I_0$, and $g$ are different for excitatory and inhibitory populations \cite{Wong2006}.

        Excitatory and inhibitory inputs to populations in the \emph{sweep layer} are modelled according to AMPA-like and GABA-like synaptic gating dynamics \cite{Brunel2001}:
        \begin{eqnarray*}
          \dot{S}_{\alpha, n}\AMPA{}(t) &=& 
                  -\frac{S_{\alpha, n}\AMPA{}(t)}{\tau\AMPA{}} + h^{\alpha e}_n(t) + \sigma \xi,
                  \quad \alpha = \uparrow, \downarrow, f
                   \\
          \dot{S}_{\alpha, n}\GABA{}(t) &=& 
                  -\frac{S_{\alpha, n}\GABA{}(t)}{\tau\GABA{}} + h^{\alpha i}_n(t) + \sigma \xi, 
                  \quad \alpha = \uparrow, \downarrow
        \end{eqnarray*}

        \noindent where $\xi$ is an uncorrelated Gaussian noise sampled independently for each synapse and instant $t$, and $\sigma = 0.0007$\,nA is the amplitude of the noise \cite{Wong2006}. The total synaptic input for each population is then:
        \begin{eqnarray*}
          I^{\uparrow e}_n(t) & = &
              J_{f}\AMPA \sum_m \omega^{f \uparrow}_{nm} S_{f, m}\AMPA{}(t - \delta t_{nm}) - \\ 
                              &   &
              J\GABA \left(\sum_m \omega^{ie}_{nm} S_{\downarrow, m}\GABA{}(t)
                                 + S_{\uparrow, n}\GABA{}(t) \right) + I\subix{bkg}^E \\
          I^{\uparrow i}_n(t)   & = & J_{s}\AMPA \sum_m \omega^{ei}_{nm} S_{\uparrow, m}\AMPA{}(t) 
                                      + I\subix{bkg}^I\\
          I^{\downarrow e}_n(t) & = &          
              J_{f}\AMPA \sum_m \omega^{f \downarrow}_{nm} S_{f, m}\AMPA{}(t - \delta t_{nm}) - \\ 
                              &   &
              J\GABA \left(\sum_m \omega^{ie}_{nm} S_{\uparrow, m}\GABA{}(t)
                                 + S_{\downarrow, n}\GABA{}(t) \right) + I\subix{bkg}^E \\
          I^{\downarrow i}_n(t) & = & J_{s}\AMPA \sum_m \omega^{ei}_{nm} S_{\downarrow, m}\AMPA{}(t)
                                      + I\subix{bkg}^I
        \end{eqnarray*}

        \noindent where $I\subix{bkg}^E$ and $I\subix{bkg}^I$ are constant background inputs putatively sourced in external neural populations \cite{Wong2006}.

        The excitatory-to-inhibitory and inhibitory-to-excitatory connectivity matrices $\omega^{ei}$ and $\omega^{ie}$ are Gaussian shaped and centred in the identity matrix:

        \begin{equation}
          \omega^{\alpha}_{nm} = e^{-\frac{(n-m)^2}{2 \sigma_{\alpha}}}, \quad \alpha = ei, ie
        \end{equation}

        \begin{figure}
          \centering
          \includegraphics[width=\columnwidth]{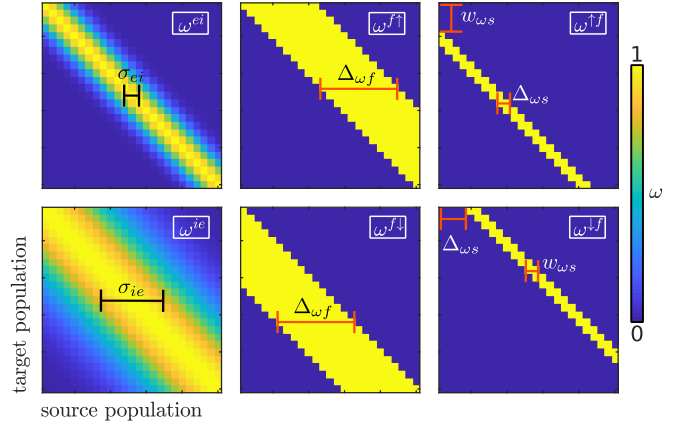}
          \caption{\textbf{Connectivity matrices.} Matrices show the connection between the first 25 ensembles of each source-target group. From left to right, matrices correspond to: excitatory-to-inhibitory $\omega^{ei}$, inhibitory-to-excitatory $\omega^{ie}$; bottom-up AMPA connections spectral-to-up $\omega^{f \uparrow}$, spectral-to-down $\omega^{f \uparrow}$; and feedback NMDA connections up-to-spectral $\omega^{\uparrow f}$, down-to-spectral $\omega^{\downarrow} f$. Labels are encircled in a white square in the top right of each plot. The free parameters of each connectivity matrix are defined geometrically in the plots.}
          \label{fig:connectivity}
        \end{figure}

        The remaining connectivity matrices $\omega^{f \uparrow}$ and $\omega^{f \downarrow}$ are defined to constraint the up (down) feed to inputs from lower (higher) frequencies and to limit the range of the connection to a small number of populations $\Delta_{\omega f}$ of the spectral representation: 

        \begin{eqnarray*}
          \omega^{f \uparrow}_{nm}   & = & \left\{
              \begin{array}{rl}
                1 & \text{ if } \quad 0 \leq n - m \leq \Delta_{\omega f}\\
                0 & \text{ otherwise}
              \end{array} \right. \\
          \omega^{f \downarrow}_{nm} & = & \left\{
              \begin{array}{rl}
                1 & \text{ if } \quad 0 \leq m - n \leq \Delta_{\omega f}\\
                0 & \text{otherwise}
              \end{array}  \right. 
        \end{eqnarray*}

        The free parameters were initialised to standard values (the effective conductivities $J_{f}\AMPA$, $J\GABA$, and $J_{s}\AMPA$, according to~\cite{Wong2006}; the baseline delay $\delta t_0$ to 2\,ms/channel; and the dispersion constants $\sigma\subix{in}$, $\sigma\subix{ei}$, $\sigma\subix{ei}$, and $\Delta_{\omega f}$, to $0.1\,N$) and manually tuned so that the networks showed direction selectivity for the FM-sweep characteristics (duration, rates, $\Delta f$) of the stimuli used in the first part of the study. Unless stated otherwise, all simulations listed in this work correspond to the parameters listed in Table~\ref{tab:pars}.

        \begin{table}
          \centering
          \caption{\textbf{Model parameters.} Most parameters were taken from the original studies that derived the mean field approximations used in the model and are cited accordingly. Other free parameters, like the number of bins of the tonotopic axis $N$, were fixed to reasonable but arbitrary values at the beginning of the model construction and were not adjusted during the analyses (\emph{ad-hoc}). Free parameters that were manually tuned are labelled as \emph{tuned (x)}, where $x$ is: 1, for parameters tuned so that the spectral layer integrates the peripheral representation correctly (see Section~\ref{sec:sw:models:spectral}); 2, for parameters tuned to achieve FM-direction selectivity; and 3, for parameters tuned so that the feedback signalling resulted in a fair fit between the model's pitch predictions and the experimental observations.}
          \begin{tabular}{cr@{\,}lc}
            parameter               &      value & (unit)         & source             \\
            $N$                     &     $100$  & channels       & ad-hoc             \\
            $dt$                    &     $0.1$  & ms             & ad-hoc             \\
            periph $dt$             &    $0.01$  & ms             & ad-hoc             \\
            periph $f_{\text{min}}$ &     $125$  & Hz             & \cite{Zilany2009}  \\
            periph $f_{\text{max}}$ &   $10000$  & Hz             & \cite{Zilany2009}  \\
            $\tau\supix{memb}_e$    &      $20$  & ms             & \cite{McCormick1985} \\
            $\tau\supix{memb}_i$    &      $10$  & ms             & \cite{McCormick1985} \\
            $\Delta$                &       $1$  & mV             & \cite{Ostojic2011} \\
            $c\supix{excitatory}$   &     $310$  & (V\,nC)$^-1$   & \cite{Wong2006}    \\
            $I_0\supix{excitatory}$ &     $125$  & Hz             & \cite{Wong2006}    \\
            $g\supix{excitatory}$   &    $0.16$  & s              & \cite{Wong2006}    \\
            $c\supix{inhibitory}$   &     $615$  & (V\,nC)$^-1$   & \cite{Wong2006}    \\
            $I_0\supix{inhibitory}$ &     $177$  & Hz             & \cite{Wong2006}    \\
            $g\supix{inhibitory}$   &   $0.087$  & s              & \cite{Wong2006}    \\  
            $I\subix{bkg}^E$        &    $0.23$  & nA             & \cite{Wong2006}    \\ 
            $I\subix{bkg}^I$        &    $0.10$  & nA             & \cite{Wong2006}    \\ 
            $\sigma$                &  $0.0007$  & nA             & \cite{Wong2006}    \\ 
            $\gamma$                &   $0.641$  &                & \cite{Brunel2001}  \\ 
            $\tau\AMPA$             &       $2$  & ms             & \cite{Brunel2001}  \\ 
            $\tau\GABA$             &      $ 5$  & ms             & \cite{Brunel2001}  \\ 
            $\tau\NMDA$             &     $100$  & ms             & \cite{Brunel2001}  \\ 
            $J\AMPA\subix{in}$      &    $0.38$  & nC             & tuned (1)          \\ 
            $J\AMPA_f$              &    $0.55$  & nC             & tuned (2)          \\ 
            $J\AMPA_s$              &    $0.67$  & nC             & tuned (2)          \\ 
            $J\GABA$                &    $0.30$  & nC             & tuned (2)          \\ 
            $J\NMDA$                &    $0.04$  & nC             & tuned (3)          \\ 
            $\sigma\subix{in}$      &   $0.1\,N$ & channels       & tuned (2)          \\ 
            $\sigma\subix{ie}$      &   $0.5\,N$ & channels       & tuned (2)          \\ 
            $\sigma\subix{ei}$      &  $0.03\,N$ & channels       & tuned (2)          \\ 
            $\Delta t_0$            &        $3$ & ms$/$channel   & tuned (2)          \\ 
            $\Delta_{\omega f}$     &  $0.05\,N$ & channels       & tuned (2)          \\ 
            $\Delta_{\omega s}$     &  $0.05\,N$ & channels       & tuned (3)          \\ 
            $w_{\omega s}$          &  $0.03\,N$ & channels       & tuned (3)          
          \end{tabular}
          \label{tab:pars}
        \end{table}

        The direction selectivity index (DSI; e.g.,~\cite{Zhang2003}) described in the Results section was computed as the proportion of the activity elicited in a network by an up sweep minus the activity elicited in the same network by a down sweep with the same duration and frequency span: 

        \begin{equation} \label{eq:dsi}
          \text{DSI}^{\alpha} = \frac{\sum_n \int\,dt \left(\left[h^{\alpha e}_n(t)\right]_{+\Delta f} - \left[h^{\alpha e}_n(t)\right]_{-\Delta f}\right)}
                              {\sum_n \int\,dt \left( \left[h^{\alpha e}_n(t)\right]_{+\Delta f} + \left[h^{\alpha e}_n(t)\right]_{-\Delta f}\right)}
          \quad \alpha = \uparrow, \downarrow
        \end{equation}

        \noindent where $\left[h^{\alpha e}_n(t)\right]_{\Delta f}$ is the firing rate $h^{\alpha e}_n(t)$ elicited in the network by a sweep with a frequency span $\Delta f$. 

      \subsubsection{Feedback connections}

        Feedback connections from the sweep layers to the spectral layer were modelled according to NMDA-like synaptic gating dynamics with a finite rising time constant \cite{Brunel2001}.

        \begin{eqnarray*}
          \dot{S}_{\alpha, n}\NMDA{}(t) & = & - \frac{S_{\alpha, n}\NMDA{}(t)}{\tau\NMDA{}} 
                  + \sigma \xi \\
                  & & + \left(1 - S_{\alpha, n}\NMDA{}(t)\right) 
                  \gamma\, h^{\alpha e}_n(t), \quad \alpha = \uparrow, \downarrow
        \end{eqnarray*}

        \noindent with $\gamma = 0.641$. NMDA currents are added to the total synaptic input of the neurons in the spectral layer as an additional term in \eqref{eq:fInputBottomUp}:

        \begin{equation*}
          I^f_n(t) \rightarrow \hat{I}^f_n(t) = I^f_n(t) +
                     J\NMDA \sum_{\alpha = \uparrow, \downarrow} \sum_m 
                      \omega^{\alpha f}_{nm} S_{\alpha, m}\NMDA{}(t)
        \end{equation*}

        The connectivity matrices $\omega^{\alpha \uparrow}_{nm}$, $\omega^{\alpha \downarrow}_{nm}$ were chosen such that the target of the NMDA-driven activation was limited to a number of $\Delta_{\omega s}$ and leave a gap of $w_{\omega s}$ populations between the centre frequency of the source and target ensembles (see Fig~\ref{fig:connectivity}, right):

        \begin{eqnarray*}
          \omega^{\uparrow f}_{nm}   & = & \left\{
              \begin{array}{rl}
                1 & \text{ if } \quad w_{\omega s} \leq m - n \leq \Delta_{\omega s} + w_{\omega s}\\
                0 & \text{ otherwise}
              \end{array} \right. \\
          \omega^{\downarrow f}_{nm} & = & \left\{
              \begin{array}{rl}
                1 & \text{ if } \quad w_{\omega s} \leq n - m \leq \Delta_{\omega s} + w_{\omega s}\\
                0 & \text{otherwise}
              \end{array}  \right. 
        \end{eqnarray*}
        
        \noindent The gap $ w_{\omega s} > 0$ is enforced to avoid resonances between sweep-selective and spectral populations with the same centre frequency during the encoding of pure tones. The free parameters were initialised to standard values (the NMDA conductivity $J\NMDA$ to the value recommended by~\cite{Wong2006}, and the connectivity parameters $w_{\omega s}$ and $\Delta_{\omega s}$ to $0.1\,N$) and manually tuned so that the pitch predictions of the model (as computed in Eq~\eqref{eq:spectralPitch}) matched the empirical data.

  \section*{Acknowledgments}

    This research was funded by the ERC Consolidator Grant SENSOCOM 647051.

    The authors would like to thank Shih-Cheng \emph{Vincent} Chien for his enlightening suggestions during the writing of the manuscript.

  \onecolumn

  \bibliography{bib}
  \bibliographystyle{ieeetr}
  
  \renewcommand\thefigure{S\arabic{figure}}
  \setcounter{figure}{0}       
  \renewcommand\thetable{S\arabic{table}}   
  \setcounter{table}{0}    
  
  \section*{Supporting information}

    \begin{figure*}[h]
      \centering
      \includegraphics[width=0.7\textwidth]{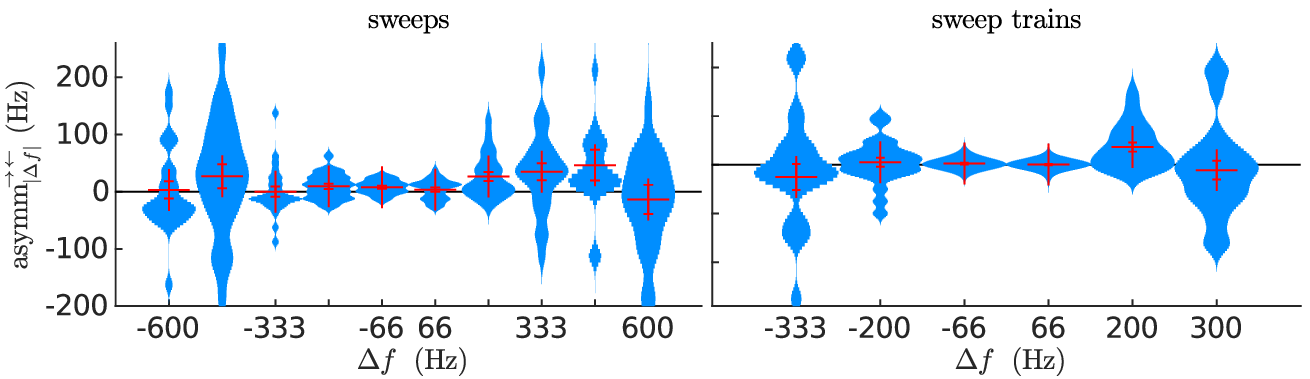}
      \caption{\textbf{Effect of the presentation order on $\Delta p$.} Kernel density estimations of the difference between the perceived pitch evaluated when the sweep was presented before the probe tone $f^{\leftarrow}_{\text{perceived}}$ and the perceived pitch evaluated when the probe tone was presented before the sweep $f^{\rightarrow}_{\text{perceived}}$; no systematic effect of the presentation order was found for any of the conditions. Each sample of the distributions corresponds to the difference of the average perceived pitch between presentation orders of the same $\Delta f$ for a given subject and centre frequency ($N = 8\times3 = 24$). Error bars show the average and the standard error of the groups.
      \label{fig:lrDev}}
    \end{figure*}

    \begin{figure*}[h!]
      \centering
      \includegraphics[width=0.7\textwidth]{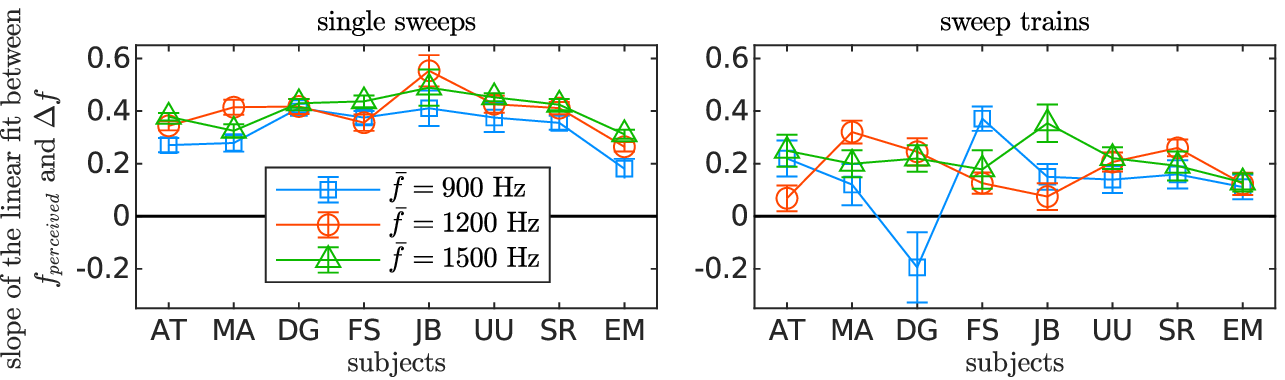}
      \caption{\textbf{Subject specific estimations of the linear fits between the pitch shift $\Delta p$ and $\Delta f$.} Each plot shows the slopes $m$ of the linear fit $f_{\text{perceived}} \sim \bar{f} + m\,\Delta f$ for each subjects, for the single sweep (left) and sweep train (right) stimuli; error bars mark the 95\% confidence intervals of the estimations. 
      \label{fig:slopes}}
    \end{figure*}    

    \begin{table}[ht!]
      \centering
      \begin{tabular}{rcccccc}
        \multicolumn{7}{c}{single sweeps (experiment I)} \\
        \hline
        $\bar{f} = $& \multicolumn{2}{c}{ 900\,Hz} & 
                      \multicolumn{2}{c}{1200\,Hz} & 
                      \multicolumn{2}{c}{1500\,Hz} 
                      \\
        probe $\leftrightarrows$ sweep & $\leftarrow$ & $\rightarrow$ & 
                      $\leftarrow$ & $\rightarrow$ & 
                      $\leftarrow$ & $\rightarrow$ \\
        \hline
        slope  ($\pm 0.03$)     &   0.36  &   0.30  &   0.38   &   0.42   &   0.41   &   0.40 \\
        $r_p$  ($p < 10^{-22}$) &   0.88  &   0.68  &   0.86   &   0.89   &   0.90   &   0.88 \\
        $r_s$  ($p < 10^{-33}$) &   0.91  &   0.78  &   0.92   &   0.96   &   0.94   &   0.96 \\    
        \hline
        & & & & & & \\
        \multicolumn{7}{c}{sweep trains (experiment II)} \\
        \hline
        $\bar{f} = $& \multicolumn{2}{c}{ 900\,Hz} & 
                      \multicolumn{2}{c}{1200\,Hz} & 
                      \multicolumn{2}{c}{1500\,Hz} 
                      \\
        probe $\leftrightarrows$ sweep & $\leftarrow$ & $\rightarrow$ & 
                      $\leftarrow$ & $\rightarrow$ & 
                      $\leftarrow$ & $\rightarrow$ \\
        \hline
        slope ($\pm 0.04$)     & $0.19$ & $0.08$ & $0.17$ & $0.18$ & $0.22$ & $0.21$ \\
        $r_p$ $(p < 10^{-7})$  & $0.54$ & n.s.   & $0.60$ & $0.61$ & $0.66$ & $0.59$ \\
        $r_s$ $(p < 10^{-3})$  & $0.55$ & $0.26$ & $0.58$ & $0.60$ & $0.66$ & $0.56$ \\
      \end{tabular}
      \caption{\textbf{Summary statistics on the relationship between the perceived pitch and the frequency span for single sweeps and sweep trains.} The slope of the linear fit, Pearson's correlation $r_p$, and Spearman's correlation $r_s$ for the relationship between $f_{\text{perceived}}$ and $\Delta f$ are presented for each centre frequency $\bar{f}$ and direction of the presentation (probe before sweep, $\rightarrow$; and sweep before probe, $\leftarrow$). Spearman's correlation is systematically larger than Pearson's, indicating that the elicited pitch is related to $\Delta f$ in a non-linear monotonic way.
      \label{tab:stats}}
    \end{table}

    \paragraph*{S1 Sounds.}
    \label{sounds}
    {\textbf{Stimuli used in the experiments.} Attached in a zipped file. Each waveform corresponds to each of the single-sweeps and sweep-trains used in the first and second experiment, respectively. File names indicate the properties of the stimulus as follows: \verb+[sweep/train]_fbar<+$\bar{f}$\verb+>_delta<+$\Delta f$\verb+>.wav+; e.g., \verb+train_fbar1200Hz_delta-333Hz.wav+ is the sweep-train with $\bar{f} = 1200\,$Hz and $\Delta f = -333\,$Hz used in the second experiment.}

\end{document}